\documentclass[twocolumn]{aastex63}

\def\lapp{\ifmmode\stackrel{<}{_{\sim}}\else$\stackrel{<}{_{\sim}}$\fi}
\def\gapp{\ifmmode\stackrel{>}{_{\sim}}\else$\stackrel{>}{_{\sim}}$\fi}
\usepackage{multirow}
\usepackage{color}
\usepackage{amsmath}
\usepackage{soul}
\usepackage{hyperref}

\shorttitle{}
\shortauthors{}
\begin{document}

\title{Investigation of the broadband emission of the gamma-ray binary HESS~J0632+057 using an intrabinary shock model}

\correspondingauthor{Hongjun An}
\email{hjan@cbnu.ac.kr}

\author{Jinyoung Kim}
\affiliation{Department of Astronomy and Space Science, Chungbuk National University, Cheongju, 28644, Republic of Korea}
\author{Hongjun An}
\affiliation{Department of Astronomy and Space Science, Chungbuk National University, Cheongju, 28644, Republic of Korea}
\author{Kaya Mori}
\affiliation{Columbia Astrophysics Laboratory, Columbia University, New York, NY 10027, USA}

\begin{abstract}
We investigated a wealth of X-ray and gamma-ray spectral energy distribution (SED)
and multi-band light curve (LC) data of the gamma-ray binary HESS~J0632+057
using a phenomenological intrabinary shock (IBS) model.
Our baseline model assumes that  the IBS is formed by colliding winds from
a putative pulsar and its Be companion, and particles accelerated in the IBS
emit broadband radiation via synchrotron (SY) and inverse-Compton upscattering (ICS) processes.
Adopting the latest orbital solution
and system geometry \citep[][]{Tokayer2021}, we reproduced the global X-ray
and TeV LC features, two broad bumps at $\phi \sim 0.3$ and $\sim0.7$,
with the SY and ICS model components.
We found these TeV LC peaks originate from ICS emission  caused by the enhanced
seed photon density near periastron and superior conjunction or Doppler-beamed
emission of bulk-accelerated particles in the IBS at inferior conjunction.
While our IBS model successfully explained most of the observed SED and LC data,
we found that phase-resolved SED data in the TeV band require an additional component
associated with ICS emission from pre-shock particles (produced by the pulsar wind).
This finding indicates a possibility of delineating the IBS  emission components
and determining the bulk Lorentz factors of the pulsar wind at certain orbital phases.

\end{abstract}


\section{Introduction}
\label{sec:intro}
High-energy $\gamma$-ray surveys using ground-based imaging  air Cherenkov telescopes 
(e.g., VERITAS, H.E.S.S. and MAGIC), along with X-ray telescopes, have uncovered a rare subclass of binary systems detected above $E\sim0.1$~TeV \citep[e.g.,][]{Corbet2012,Aharonian2006}. These so-called TeV $\gamma$-ray binaries (TGBs) harbor a compact object and a massive companion (O, B or Be star), with a wide range of orbital periods spanning 3.9~days to $\sim50$~years. TGBs emit orbitally-modulating broadband radiation from X-ray to gamma-ray energies  \citep[e.g.,][]{Mirabel2012}. 
It is widely accepted that very high-energy (VHE; $\ge$0.1\,TeV) emission 
from TGBs implies that particles should be accelerated to GeV--TeV energies in the system \citep[][]{Becker2017}. 
Among $\le$10 TGBs discovered thus far, the compact object has been identified as a neutron star in only three systems: PSR~B1259$-$63 \citep[][]{Johnston1992}, PSR~J2032+4127 \citep[][]{Ho2017}, and LS~I~+61$^\circ$~303 \citep[][]{Weng2022}.

Studies of TGBs have been carried out primarily by modeling their light curves (LCs) and broadband spectra in the X-ray and gamma-ray bands. The multi-wavelength spectral energy distributions (SEDs) are well characterized by two non-thermal components at low energy ($\le$100\,MeV) and gamma-ray bands ($\sim$TeV) which are mixed with thermal emission from the companion.
It is thought that the low-energy non-thermal emission extending from radio to $\le$100\,MeV band 
is produced by synchrotron (SY) radiation of energetic electrons \citep[e.g.,][]{Tavani1994,Dubus2013}. The GeV emission may consist of the SY and the pulsar magnetospheric radiation with some contribution from inverse-Compton scattering (ICS) of stellar thermal photons
by low-energy electrons \citep[e.g., with a Lorentz factor of $\sim10^4$;][]{Zabalza2013,Dubus2015}. 
The VHE emission is believed to be produced by ICS
of the stellar photons by high-energy electrons \citep[e.g.,][]{Dubus2006,Khangulyan2008,Chernyakova2020}. 
The X-ray and VHE emission, resulting from SY and ICS in the shocked region, respectively, shows a strong dependence on orbital phase due to various factors such as the intrabinary distance, anisotropic radiation processes and relativistic Doppler boosting. 

These emission mechanisms have been employed primarily in two scenarios for TGBs:
a microquasar and an intrabinary shock (IBS) scenario. 
In the microquasar scenario, the `unknown' compact object is assumed to be a
black hole with bipolar and relativistic jets. Particles are accelerated to high energies in the jets, and orbital variation of the jet viewing angle generates
the orbital modulation in the emission \citep[e.g.,][]{Bosch-Ramon2007,Marcote2015}.
In the IBS scenario, the compact object is assumed to be a pulsar 
whose wind interacts with the companion's
outflow. The wind-wind collision produces a contact discontinuity (CD) in the IBS region where pulsar wind particles are accelerated 
and emit broadband non-thermal radiation \citep[e.g.,][]{Dubus2006}.
Orbital variation of the orientation of the IBS flow with respect to the observer’s line of sight (LoS) causes the high-energy emission to modulate with the orbital period. \citep[e.g.,][]{vandermerwe2020}. 
Given the three TGBs containing radio pulsars combined with constraints on mass function, the compact object in other TGBs is generally considered to be a neutron star \citep[e.g.,][]{Dubus2013}. 

The TeV source HESS~J0632+057 (J0632 hereafter) was identified as a TGB by a detection of its $\sim320$-day orbital modulation in the VHE band \citep[][]{Acciari2009}. Later, X-ray and GeV modulations on the orbital period ($P_{\rm orb}$) were detected \citep[][]{Aliu2014,Li2017}, confirming the VHE identification. 
An optical spectroscopic study identified the companion to be a Be star \citep[HD~259440;][]{Aragona2010}
with an equatorial disk. The compact object has not been identified yet, but
a Chandra imaging found a hint of extended emission
around the source which was interpreted as a signature of wind-wind
interaction \citep[][]{Kargaltsev2022}.

In this paper, we test the orbital solution suggested
by TAH21 and determine the IBS parameters 
of J0632 using the $\sim$GeV and VHE measurements. Our phenomenological model fit to the extensive X-ray and gamma-ray data puts some constraints on magnetohydrodynamic (MHD) flows in the IBS that are useful to further MHD simulations of J0632 and other TGBs. 
We describe IBS structure in Section~\ref{sec:sec2} and present emission model components in
Section~\ref{sec:sec3}. We then use the model to explain the broadband data of J0632
in Section~\ref{sec:sec4}. We discuss implications of the modeling in Section~\ref{sec:discussion}
and present a summary in Section~\ref{sec:summary}.

\section{orbital solutions for J0632}\label{sec2_0}
\begin{figure}
\centering
\includegraphics[width=80mm]{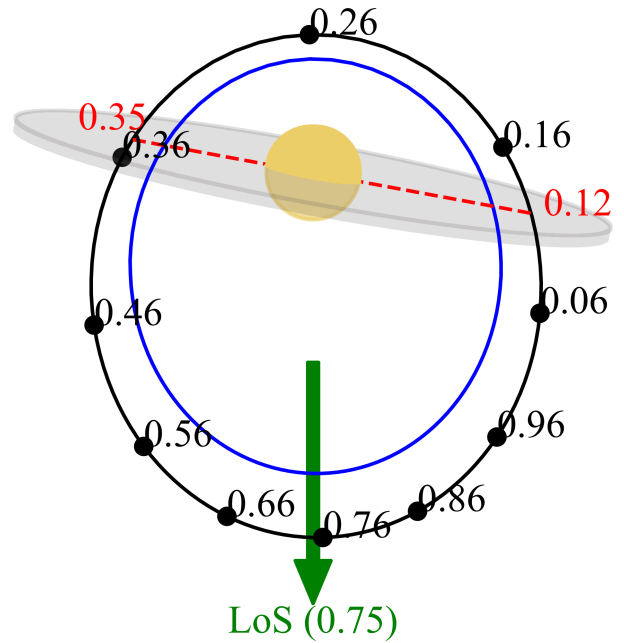} \\
\figcaption{An orbit of the J0632 system suggested by \citet{Tokayer2021}.
The black dots represent pulsar's position at various orbital phases,
the yellow circle is the companion (Be star; not to scale),
its disk is depicted by a gray disk, and the blue circle shows the locus of the shock apex.
Pulsar crossing phases are denoted in red. The green arrow shows the direction towards the observer.
See Figure~8 of \citet{Tokayer2021} for other suggested orbits of J0632.
\label{fig:fig1}}
\end{figure}

The X-ray and the VHE LCs of J0632 exhibit a similar shape characterized by  broad bumps at
orbital phases $\phi\approx0.25$ and 0.75, and a sharp spike at $\phi$=0.35 \citep[e.g.,][]{Tokayer2021, Adams2021}. 
These features could probe the emission  mechanisms 
and help infer the properties of particle acceleration and flow in the binary system. However, the orbital solution of J0632 has not been well determined.
While optical data have provided
accurate orbital solutions for other TGBs, the situation for J0632 is unclear.
Orbital solutions derived from radial
velocity measurements using the H$\alpha$ line do not agree
with each other. 
For the reference epoch of MJD~54857 used throughout this paper,
the solution inferred from an optical study by \citet{Casares2012} suggests
a highly eccentric orbit with an eccentricity of 0.83,
periastron at $\phi=0.967$, and LoS at $\phi=0.961$. In contrast, \citet{Moritani2015} suggested
a less eccentric orbit with eccentricity of 0.64, periastron at $\phi=0.663$, and LoS at $\phi\approx 0.17$.
Even considering that their radial velocity curves were folded on different periods,
321\,day and 313\,day in \citet{Casares2012} and \citet{Moritani2015}, respectively,
the solutions differ substantially.

X-ray LC data provide alternate orbital solutions to those obtained with optical data.
Two such solutions are obtained by attributing the bumps in the X-ray LC to disk interactions \citep{Malyshev2019,Chen2022}.
Most recently, \citet{Tokayer2021} derived an orbit of J0632 by modeling the most extensive X-ray LC data 
with an IBS model (Fig.~\ref{fig:fig1}).
This latest orbital solution seems to be well justified since the model matches the X-ray LC data well
and accounts for  the enhanced hydrogen column density ($N_{\rm H}$) observed
at some orbital phases which \citet{Malyshev2019} and \citet{Tokayer2021}
attributed to the pulsar-disk interaction.
These three orbits inferred from modelings of X-ray LCs folded on $P_{\rm orb}\approx317$\,day
are all similar to one another even though
the emission models employed in those works are somewhat different. The suggested
eccentricities are modest (0.4--0.5), and phases of periastron and LoS are $\phi=$0.3--0.4 and 0.7--0.8,
respectively.
Considering that the three X-ray-inferred orbits broadly agree, this leaves three orbital solutions 
for J0632 which significantly disagree with one another -- two from optical data and one from X-ray studies.
Because it is unclear which orbital solution is correct,
it will be helpful to check to see if the suggested
orbits can explain VHE data (LC and SED) using emission models, which has not been
done for any of the suggested orbits.

The optical orbits of \citet{Casares2012} and \citet{Moritani2015} are incompatible with
a shock-emission scenario as noted by \citet{Chen2022}. Thus they are inadequate for
an IBS study of J0632.
The X-ray orbits of \citet{Chen2022} and \citet{Malyshev2019}
were constructed using an inclined disk model which employs
a termination shock and its interaction with an inclined disk. This is essentially a one-zone
model with a disk as the shock region is assumed to be a point source \citep[][]{Chen2022},
whereas the IBS model of \citet[][TAH21 hereafter]{Tokayer2021}
took into account the multi-zone emission from a cone-shape IBS region.
Because the IBS, unlike the one-zone termination shock, produces Doppler-boosted emission
along the shock tail \citep[e.g.,][]{Dubus2013, An2017, vandermerwe2020},
the orbit of J0632 inferred by an IBS model (Fig.~\ref{fig:fig1}; TAH21)
slightly differs from that obtained by the inclined disk model.
Note that TAH21 also applied a one-zone shock model to the broadband SEDs of J0632,
but they did not attempt to model the VHE LC data. 
Given that TGBs should form an extended IBS as demonstrated
by hydrodynamic simulations \citep[e.g.,][]{Bogovalov2008,Dubus2015,Bosch-Ramon2017}, 
it is necessary to consider a multi-zone shock case for modeling both the multi-wavelength SED and LC data in detail.
Hence, we restrict our study to an IBS scenario and the orbit of TAH21;
the distinct features in the X-ray and VHE LCs allow us to determine the IBS properties more accurately.

\section{Structure of IBS}\label{sec:sec2}
\begin{figure*}
\centering
\begin{tabular}{ccc}
\hspace{-10 mm}
\includegraphics[width=75mm]{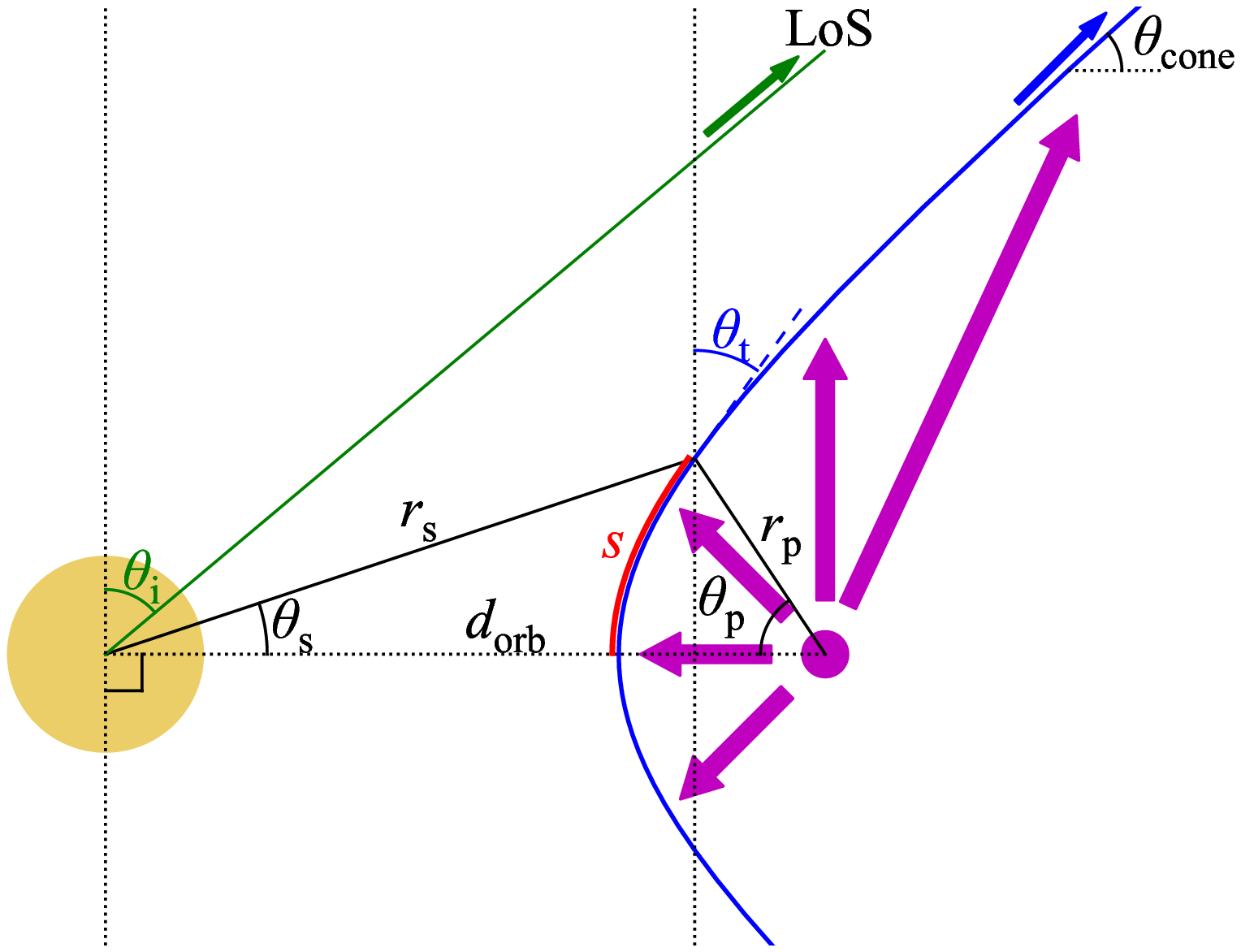} &
\includegraphics[width=75mm]{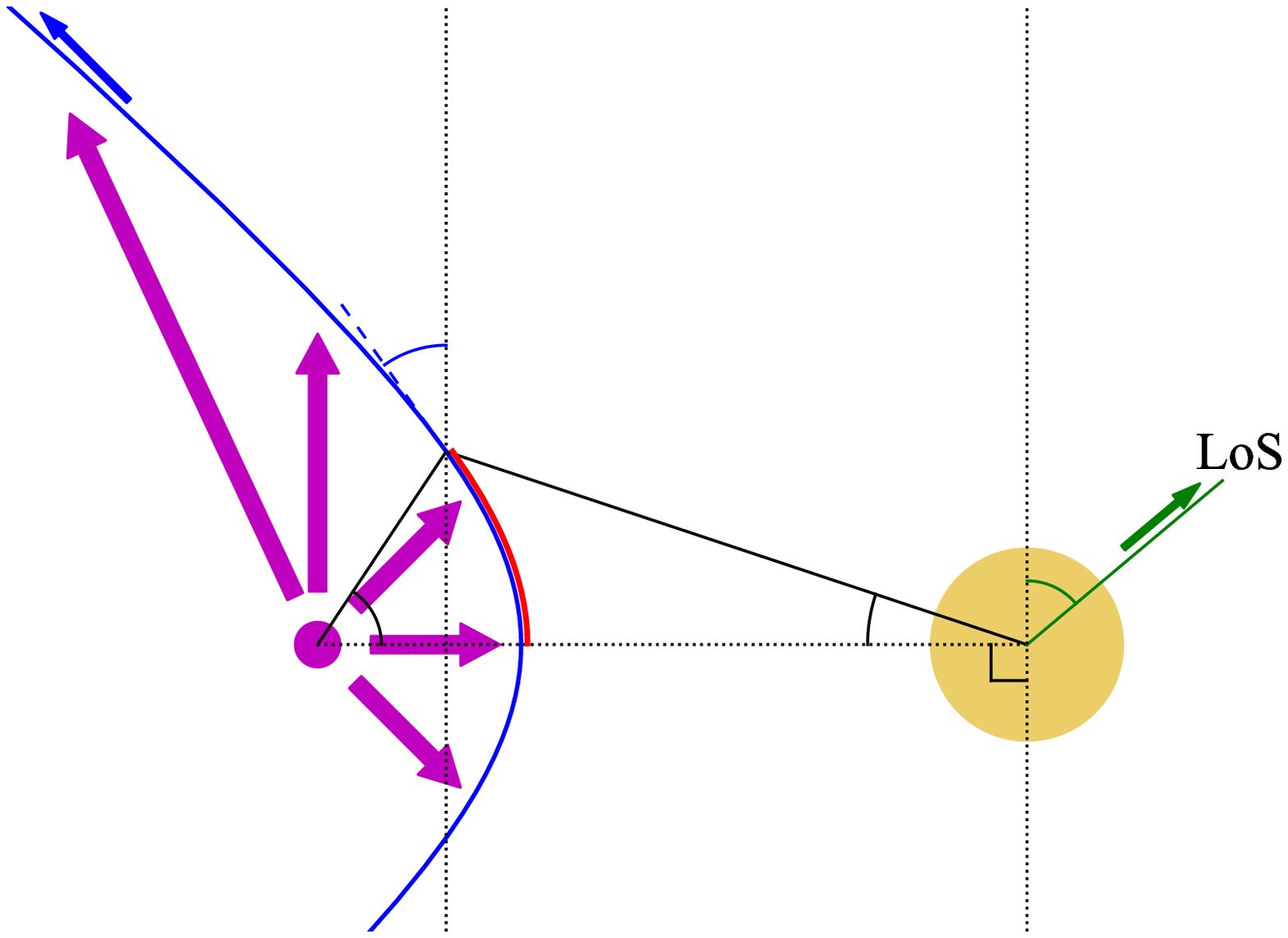} \\
\end{tabular}
\figcaption{Vertical cross sections of a TGB system at the inferior (left)
and superior conjunction (right) of the pulsar. The observer's line of sight (LoS) is in the right direction.
The companion star is shown in yellow (not to scale),
a pulsar at an orbital separation $d_{\rm orb}$ is denoted by a purple dot,
and CD is drawn as a blue curve. Arc length along the CD to an emission zone from the shock nose is $s$ (red).
Angles to the emission zone from the line of centers are
$\theta_{\rm p}$ (from the pulsar) and $\theta_{\rm s}$,
and distances to the zone from the pulsar and the companion are $r_{\rm p}$ and $r_{\rm s}$, respectively.
$\theta_{\rm t}$ and $\theta_i$ are polar angles
of the local flow tangent (blue dashed) and the observer inclination (green), respectively.
The IBS half opening angle is denoted as $\theta_{\rm cone}$.
Purple arrows show flow of the preshock pulsar wind.
\label{fig:fig2}}
\end{figure*}

In our IBS model, a pulsar injects 
cold pulsar-wind (preshock; purple arrows in Fig.~\ref{fig:fig2})
electrons and magnetic field ($B$) to the IBS (blue line in Fig.~\ref{fig:fig2}),
and the electrons are accelerated to very high energies in the shock 
\citep[e.g.,][]{Tavani1997,Dubus2015}. These particle injection and acceleration schemes have been widely used in the previous models
of TGBs \citep[e.g.,][]{Sierpowska-Bartosik2008,Dubus2015,An2017,Chernyakova2020,Xingxing2020}.
In this work, we adopt an analytical approach for modeling the global features of the flows, although hydrodynamics (HD) simulations showed more complexities and substructures in the particle flows \citep[e.g.,][]{Bosch-Ramon2015}.  Below we describe the prescriptions, assumptions, and formulas for the IBS flows in our model.

\subsection{Pulsar wind and stellar outflow}\label{sec:sec2_1}
	The 2D shape of an IBS is determined by the pressure balance of the pulsar and the stellar winds.
The pulsar wind is thought to be composed of cold and relativistic plasma.
MHD simulations suggest that the pulsar wind can be anisotropic with higher flow velocities and particle densities in the equatorial plane \citep[e.g.,][]{Tchekhovskoy2016}.
Massive stars emit isotropic outflows and Be-type stars such as the companion star of J0632 may have strong equatorial outflows 
(decretion disk; e.g., Fig.~\ref{fig:fig1}) which
are evidenced by an infrared (IR) excess  \citep[e.g.,][]{Waters1984}.
Besides, stellar outflows can be clumpy and highly variable,
inducing large variability to the IBS shape and thus to the observed
emission of TGBs \citep[e.g.,][]{Bosch-Ramon2013}.

	The shape of an IBS formed by anisotropic wind interaction is difficult to
determine observationally given that the pulsar's spin axis and anistoropic geometry of the pulsar and/or stellar winds are not known. 
In this work, we assume an isotropic geometry for both the pulsar and the stellar winds
since an IBS formed by slightly anisotropic winds does not differ much from the isotropic case  \citep[e.g.,][]{Kandel2019};
the most important shock tangent angle near the tail
changes only slightly, which can be accounted for by a small change
in the wind momentum flux ratio and/or observer's inclination angle in
our phenomenological model. Note, however, that the strong equatorial outflow
(i.e., disk) of a Be-type companion can significantly alter the IBS shape
at the pulsar crossing phases (e.g., phases 0.12 and 0.35 in Fig.~\ref{fig:fig1}).

Under the isotropic wind assumption, our model does not properly account for the IBS-disk interaction
at the disk-crossing phases. Moreover, emission produced by the interaction
depends strongly on other parameters such as density and heating
that are poorly known, and need to be determined by simulations and observations of the interaction sites.
We do not consider such
interaction in this work and hence our model does not naturally reproduce the
X-ray and TeV spikes at $\phi\approx0.3$. Note, however, that TAH21 arbitrarily
increased $B$ at the interaction phases to reproduce the X-ray spike. We adopt this
prescription for the X-ray modeling and thus our model phenomenologically matches
the X-ray LC, but not the VHE LC (see Section~\ref{sec:sec4_2_1} for more details).

\subsection{Shape of the IBS}\label{sec:sec2_2}
The orbital variability of high-energy emission observed in TGBs is thought to be induced
by a change in the emission and viewing geometry of the IBS \citep[e.g.,][]{Romani2016}.
In general, the IBS is assumed to have a paraboloid shape near the apex but it is distorted significantly
at large distances from the system  \citep[e.g.,][]{Bosch-Ramon2017}.
Because the high-energy emission in the X-ray to VHE band
is mostly produced in the inner regions of the IBS \citep[e.g.,][]{Dubus2015,Bosch-Ramon2017}, 
the analytic formulas (Appendix~\ref{sec:appendix1}) presented in \citet{Canto1996}
for CD produced by isotropic wind-wind
interaction is adequate for our modeling effort.
The IBS shape is basically determined by the winds' momentum flux ratio:
\begin{equation}
\label{eq:beta}
\beta=\frac{\dot E_{\rm SD}}{\dot M_{\rm w}v_{\rm w}c},
\end{equation}
where $\dot E_{\rm SD}$ is the pulsar spin-down power, $\dot M_{\rm w}$ is the mass loss rate of
the companion, $v_{\rm w}$ is the velocity of the companion's wind, and $c$ is the speed of light.
The massive companion's wind is likely stronger than the pulsar's ($\beta<1$) thus it is likely that 
the IBS is formed around the pulsar in TGBs.

A schematic view of the vertical cross sections of a TGB system
and its IBS is depicted in Figure~\ref{fig:fig2}.
As denoted in the figure, an emission zone at a distance $s$ (red solid line) along the IBS (blue solid)
from its apex is $r_{\rm s}$ away from the star at an angle $\theta_{\rm s}$
and is $r_{\rm p}$ away from the pulsar at an angle $\theta_{\rm p}$.
The blue dashed line shows the direction of the particle flow in the emission zone (polar angle $\theta_{\rm t}$),
and the green line is an observer's LoS with an inclination $\theta_i$.
The asymptotic tangent angle (blue arrow; half opening angle $\theta_{\rm cone}$) of the shock
is determined by $\beta$ (Eq.~\ref{eq:coneangle}).
These geometrical parameters for an IBS are computed using the equations
given in Appendix~\ref{sec:appendix1} \citep[see][for more detail]{Canto1996}.

	Both $r_{\rm s}$ and $r_{\rm p}$ vary orbitally because they are proportional to 
the orbital separation ($d_{\rm orb}$) between
the pulsar and the companion (e.g., Eq.~\ref{eq:shocknose}):
\begin{equation}
\label{eq:dorb}
d_{\rm orb}(\phi_0)=\frac{a(1-e^2)}{(1+e\mathrm{cos}\phi_0)},
\end{equation}
where $a$ is the semi-major axis, $e$ is the eccentricity, and $\phi_0$ is the true anomaly.
We assume that the emission zone extends to $s_{\rm max}=3$--$5 d_{\rm orb}$
\citep[e.g.,][]{Dubus2015,Bosch-Ramon2017}.
Note that the emission-zone size ($s_{\rm max}$) in our model
is assumed to be a constant multiple of $d_{\rm orb}$ which changes
with the orbital phase in eccentric orbits (Eq.~\ref{eq:dorb}).
However, the solid angle subtended by the IBS from the pulsar
(i.e., pulsar's energy injection into IBS) remains unchanged.

\subsection{Particles in the pulsar-wind zone}\label{sec:sec2_3}
	The preshock pulsar wind is composed of cold electrons accelerated
in the pulsar wind zone (purple arrows in Fig.~\ref{fig:fig2}).
The exact location and physical processes of the
particle acceleration are not well known, and hence
the energy distribution of the particles is unclear.
Some physical processes may produce narrow Maxwellian-like distributions \citep[e.g.,][]{Hoshino1992,Sironi2011}
while others may produce a broad power-law distribution \citep[e.g.,][]{Jaroschek2008}.
As such, various distributions have been assumed in TGB emission models previously:
a broadened delta function \citep[e.g.,][]{Khangulyan2011},
power-law \citep[][]{Sierpowska-Bartosik2008}, or Maxwellian distribution \citep[][]{Takata2017}.
These distributions would result in slightly different shapes for the ICS SED of
the preshock particles.
In this work, we assume that the preshock particles
are accelerated near the light cylinder $R_{\rm LC}\ll r_{\rm p}$ of the pulsar \citep[e.g.,][]{Aharonian2012},
flow isotropically, and follow a relativistic Maxwellian energy distribution
\begin{equation}
\label{eq:maxwellian}
\frac{d\dot N_e^{\rm pre}}{d\gamma_e}=N_0 \frac{\gamma_e^2\beta_e}{\Theta K_2(1/\Theta)}e^{-\gamma_e/\Theta},
\end{equation}
where $\beta_e$ is $\sqrt{1 - 1/\gamma_e^2}$, $K_2$ is the modified Bessel function of the
second kind, and $\Theta$ is a temperature parameter which we adjust to have
a Lorentz factor $\gamma_{\rm e,peak}^{\rm pre}\approx 10^6$ at the peak
of the distribution \citep[e.g.,][]{Amato2021}.
The number and energy of the particles injected into the preshock by the pulsar are given by
\begin{equation}
\label{eq:ndot}
\dot N=\int \frac{d\dot N^{pre}_e}{d\gamma_e} d\gamma_e,
\end{equation}
and
\begin{equation}
\label{eq:number}
\int \gamma_e m_e c^2 \frac{d\dot N^{pre}_e}{d\gamma_e} d\gamma_e = \eta \dot E_{\rm SD},
\end{equation}
where $\eta$ is the particle conversion efficiency of the pulsar's
spin-down power \citep[e.g.,][]{Gelfand2009, Uchiyama2009}.
Then, the number of particles within a radial length $dr$ over the $4\pi$ solid angle in the pulsar wind zone
is given by $\frac{dN}{dr}=\frac{\dot N}{c}$.
At certain orbital phases, the preshock flow along the LoS may be open if it
does not cross the IBS. In this case, we stop the flow at a distance $\approx 5d_{\rm orb}$
where a back shock is expected to form \citep[][]{Dubus2015}. We verified that extending the
preshock flow to larger distances at these phases had no significant impact on the output emission.

\subsection{Particles and their flows in IBS}\label{sec:sec2_4}
	The preshock particles are injected into the entire surface of the IBS
and further accelerated there as in termination shocks of
pulsar wind nebulae \citep[PWNe;][]{Kennel1984}.
The particles flow along the conic surface (blue in Fig.~\ref{fig:fig2})
towards the tail of the IBS and exit the emission zone at $s=s_{\rm max}$.
As IBS and PWN physics share some common grounds, we assume that the IBS electron distribution
in the flow rest frame is isotropic
and follows a broken power law \citep[e.g., as in PWN cases;][]{Sironi2011,Cerutti2020}
between a lower ($\gamma_{e,\rm min}$) and an upper ($\gamma_{e,\rm max}$) bound with
a break at $\gamma_b$ which is caused by particle cooling:
\begin{eqnarray}
\label{eq:bpl}
\frac{dN_e}{d\gamma_e} = 
\begin{cases}
N_1 \gamma_e^{-p_1}, & \gamma_{e,\rm min}\le \gamma_e<\gamma_b \\
N_1 \gamma_b^{-p_1}(\gamma_e/\gamma_b)^{-p_2}, & \gamma_b\le\gamma_e \le \gamma_{e,\rm max}. \\
\end{cases}
\end{eqnarray}
$p_1$ may be determined by particle acceleration modeling or more directly by X-ray observations.
It is almost certain that $p_1$ varies over the orbit since gamma-ray binaries have shown orbital variations of their X-ray spectral index \citep[][]{Bosch-Ramon2005,Chernyakova2009,Corbet2012,An2015}.
Because the origin of the variability is still uncertain and yet 
it does not have a significant impact on the LCs,
we assume a constant $p_1$ over the orbit.

	In our model, the particle cooling is manifested as a spectral break
of the stationary electron distribution (Eq.~\ref{eq:bpl}).
In the case that $B$ is uniform over the IBS, a $p_2 - p_1=1$ break is
expected. However, the degree of the spectral break ($p_2 - p_1$) may differ in IBSs because $B$ varies with $s$ and fresh particles are injected at every point on the IBS.
Cooling time scale of the highest-energy electron is an order of $\le$100\,s
(e.g., $\gamma_{e,\rm max}\approx 10^8$ and $B\approx1$\,G)
and then a spectral break is expected at
$\gamma_b \sim 5\times 10^{6}$. For typical $B\approx 1$\,G in IBSs of TGBs, particles
at the break would emit $\sim$MeV photons;
$p_2$ and $\gamma_b$ cannot be inferred directly from data
due to the lack of sensitive observations in the MeV band.
For $\gamma_{e,\rm max}$, we assume that particles are accelerated to the radiation reaction limit
so that the SY photon energy emitted by the highest-energy particles is $\sim 100$\,MeV
(but see Section~\ref{sec:sec4_2_1}).

	The flow bulk Lorentz factor $\Gamma$ can vary along the IBS in a complex way
as was seen in relativistic HD simulations \citep[e.g.,][]{Bogovalov2008,Dubus2015},
where a small fraction $\xi$ ($\ll 1$) of the particles in the flow is seen to be bulk-accelerated
to high $\Gamma$. 
We follow \citet{An2017} and TAH21, and for simplicity
assume two distinct populations of particles in the IBS,
one with a small constant bulk Lorentz factor ($\Gamma\approx 1$; slow flow) and
another with a linearly growing bulk Lorentz factor
\begin{equation}
\label{eq:bulkG}
\Gamma(s) = 1 + \frac{s}{s_{\rm max}}(\Gamma_{\rm max} - 1)
\end{equation}
(fast flow).
The flow speed ($v_{\rm flow}=c\sqrt{1-1/\Gamma^2}$) of the `fast flow'
is determined by this equation, but that of the slow flow
needs to be prescribed. In an ideal model for PWNe \citep[][]{Kennel1984}, the flow speed in
the post-shock region is predicted to be $\approx c/3$, and we use a similar value
for the slow flow (but see below).
Flows with different velocities are subject to instabilities
and thus may not be stable. Our phenomenological description of the two flows
does not account for the physical instabilities, and we assume that the two flows are physically separated.
More accurate description of the flow requires relativistic HD simulations as noted above;
the speed change from the `slow' to `fast' flow may be more continuous \citep[e.g.,][]{Bogovalov2008,Dubus2015}.

	The number of particles per unit length along the IBS
(integrated over the azimuth angle $\phi_s$ of the shock cone)
is given by the continuity condition \citep[e.g.,][]{Canto1996} as
\begin{equation}
\label{eq:IBSparticles}
\frac{dN_e(s)}{ds} = \frac{\dot N [1 - \mathrm{cos}\theta_{\rm p}(s)]}{2v_{\rm flow}},
\end{equation}
where $\dot N$ is the number of preshock particles injected by the pulsar given in Eq.~\ref{eq:ndot}. The number of particles in the IBS $N_e = \int \frac{dN_e(s)}{ds} ds$ is controlled by particle residence time
$s_{\rm max}/v_{\rm flow}$. 
This determines the
relative contribution of the preshock and the IBS particles to ICS (VHE) emission.
As the pulsar's injection into the IBS is assumed to be the same at every orbital phase,
we assume that $N_e$ in the IBS is constant over the orbit. This means that $v_{\rm flow}$ varies over the orbit; once this value
is prescribed at a given phase,
values at the other phases are determined to preserve $N_e$. Note that this assumption
does not have a large impact on matching the observational data with the phenomenological model;
using a different assumption (e.g., varying $N_e$ and constant $v_{\rm flow}$)
can also reproduce the measurements with slightly different $\theta_i$ and $\Gamma_{\rm max}$.

We split the IBS cone, seen in Figure~\ref{fig:fig2}, into $21\times361$ computational grids which are described in 
greater detail in Sections~\ref{sec:sec3_2}, \ref{sec:sec3_3}, and \ref{sec:sec4_2}. We apply
the same particle distribution over the IBS zones (Eq.~\ref{eq:bpl}), but the number of particles (Eq.~\ref{eq:IBSparticles}),
the fast flow $\Gamma$ (Eq.~\ref{eq:bulkG}), and the flow direction (Fig.~\ref{fig:fig2}) differ
in each zone as they depend on $s$.

\subsection{Magnetic field in IBS}\label{sec:sec_2_5}
The magnetic field in an IBS is assumed to be supplied by the pulsar and randomly oriented.
For a pulsar with a surface magnetic-field strength $B_{\rm s}$ and a spin period $P_{\rm s}$,
$B(s)$ in an emission zone in the IBS is estimated to be
\begin{equation}
\label{eq:Bfield}
B(s)=B_{\rm s}\left (\frac{2\pi R_{\rm NS}}{cP_s}\right )^3
\left( \frac{cP_{\rm s}}{2\pi r_{\rm p}(s)} \right )\equiv B_0\left ( \frac{r_0}{r_{\rm p}(s)} \right ),
\end{equation}
where $R_{\rm NS}$ is the radius of the neutron star and $r_0$ is distance to a reference point.
The pulsar continuously injects particles and $B$ which are frozen together as they stream to the IBS, 
potentially creating complex $B$ structures within the IBS flow.
Relativistic HD computations \citep[e.g.,][]{Bogovalov2012,Dubus2015} indeed
show very complicated structures of $B$: nonlinear changes with $s$, variations over the thickness of the IBS, and most importantly a $B$ that decreases with increasing $s$.
However, the simplified binary pulsar magnetic relationship in Eq.~\ref{eq:Bfield} is an appropriate substitute for the 
complex MHD results as it captures the important inverse $B$-$s$ relationship.  Prior analyses of IBSs of pulsar binaries and TGBs have used this approach 
\citep[e.g.,][]{Romani2016,Dubus2013}, and we adopt Eq.~\ref{eq:Bfield} in this work as well.
In reality, deviations from the $B\propto 1/r_{\rm p}$ dependence could be
present, but an investigation of such effects is beyond the scope of this paper
which attempts to account for the global IBS features and SED/LC data.

The value of $B_0$ can vary substantially depending on the pulsar parameters
and is often assumed to be $\approx 1$\,G in TGBs \citep[e.g.,][]{Dubus2013}.
In our model, $B_0$ at the reference point (i.e., the shock nose at the inferior conjunction of the pulsar)
is prescribed, and $B$s at the other positions in the IBS and at different orbital phases
are computed using Eq.~\ref{eq:Bfield}.

\section{Emission in TGB systems}\label{sec:sec3}
	The observed emission of TGB systems spans over the entire electromagnetic wavelengths from radio to TeV gamma-ray bands.
The radio-emission region is often observed to be elongated \citep[e.g., $>d_{\rm orb}$;][]{Ribo2008,Moldon2011}
and so the radio emission is thought to be produced
by particles that escaped from the IBS which we do not model.
The equatorial disk of a Be companion emits in the IR to optical band,
and the massive companion (blackbody) radiates
primarily in the optical band. These IR and optical photons provide
the preshock and IBS particles with seeds for ICS.

	Below, we describe SY and ICS emissions of the preshock and IBS particles relevant
to X-ray and VHE emissions in detail. IR and optical photons from the disk and the companion play an  important role for the VHE emission via ICS.

\subsection{Stellar emissions}\label{sec:sec3_1}
\begin{figure}
\centering
\includegraphics[width=80mm]{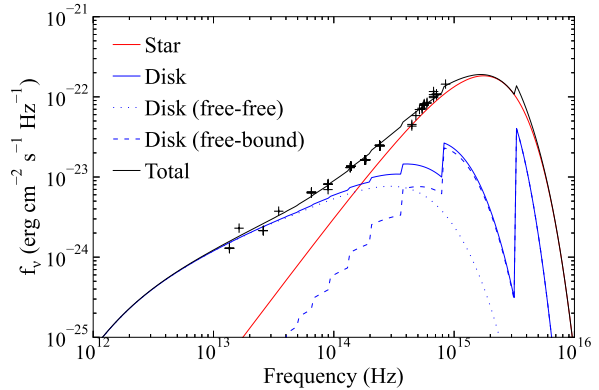} \\
\figcaption{Measured IR-to-optical flux densities of the companion in J0632
and a stellar+disk emission model.
The stellar blackbody model is displayed in a red solid line, the disk emission model
is shown by a blue solid line, and the summed model is presented as a black solid curve.
Free-free and free-bound emissions of
the disk are also plotted in blue dotted and dashed lines, respectively.
The data are taken from the {\tt VizieR} photometry viewer,
and the extinction is corrected with $A_V=2.4$ \citep[][]{ZhuHui2017}.
\label{fig:fig3}}
\end{figure}

    Seed photons for ICS are primarily  produced in the atmosphere
and the disk of the companion star.
The atmospheric emission is presumably isotropic blackbody radiation with
absorption/emission lines and is observed as a prominent optical bump
in SEDs of TGBs. This emission can be well characterized by the temperature
$T_{*}$ and radius $R_{*}$ of the star that can be measured with spectroscopic observations.
Although there are a number of absorption/emission lines in the spectrum of a star, these narrow features
will be washed out in the ICS processes by electrons with featureless, non-thermal energy distributions.
Hence, a blackbody spectrum, scaled to the distance to
an emission zone ($r_{\rm s}$; Fig.~\ref{fig:fig2}), is a good approximation for ICS seed photons.
A blackbody spectrum of J0632 computed with $T_*=30000$\,K and $R_*=6.6R_{\odot}$ \citep[][]{Aragona2010}
is presented in red in Figure~\ref{fig:fig3} along with
observed J0632 IR-to-optical flux densities.\footnote{http://vizier.unistra.fr/vizier/sed/}

	Disk emission of a Be star is produced by
free-free and free-bound processes in the disk \citep[e.g.,][]{Waters1984,Carciofi2006}
and is observed as an IR excess in the spectrum \citep[e.g.,][]{Klement2017}. While  the observed
disk spectra appear simple (e.g., broad IR bump; Fig.~\ref{fig:fig3}), computing the emission spectrum
is highly complicated because it is not uniform over the surface of the disk
due to changes of the disk density and thickness with distance from the star. 
Moreover, the disk emission is anisotropic \citep[e.g.,][]{Carciofi2006} because of
varying optical depth depending on the disk viewing angle ($\theta_{\rm d}$).

	The disk emission itself is usually much weaker than the stellar atmospheric one
\citep[e.g., Fig.~\ref{fig:fig3}; see also][]{Klement2017} and therefore EC emission off of disk seed photons
does not contribute much to VHE emission in
TGBs \citep[e.g.,][see also Section~\ref{sec:sec3_3}]{vanSoelen2012}. Furthermore,
a detailed shape of the disk spectrum is blurred in the ICS process by the
broad electron distributions. Hence, it is well justified to use an approximate
continuum model for the disk emission (e.g., multi-temperature blackbody), but
we investigate a more complex disk-emission model below.

	Although a complete disk model (considering radiation transfer
and radiative equilibrium) requires computations of the disk structure and its emission
for arbitrary density and velocity distributions \citep[e.g.,][]{Carciofi2006},
the disk emission can be simplified for our purpose with an assumption of power laws
for radial distributions of density,
scale height, and temperature.
In a cylindrical coordinate system for the disk, \citet{Carciofi2006} assumed
$\rho(r)=\rho_0( R_*/r )^{n_d}\mathrm{exp}\left (-z^2/2H^2 \right )$, $H(r)=H_0 (r/R_*)^{\beta_d}$, and
$T_{\rm disk}(r)=T_d(r/R_*)^{-s_d}$, for the density, scale height, and temperature,
respectively, and presented analytic formulas for computation of optical depths ($\tau(r)$) and
emission spectra of pole-on disks.
$n_d$ and $\beta_{\rm d}$
are typically $\sim$3 and $\sim$1.5, but they vary from source to source \citep[][]{Klement2017}.
We followed the procedure described in the Appendix of \citet{Carciofi2006}
for computations of optical depths and the disk spectra. 
Note that these computations require Gaunt factors for which \citet{Carciofi2006} used
a long-wavelength approximation. 
We instead used approximations from \citet{Waters1984}, which are more accurate within our frequency range of $10^{12}$--$10^{16}$\,Hz.
For an inclined disk as compared to a pole-on disk, the depth of the emitting medium (disk)
increases by a factor of $1/\mathrm{cos}\theta_{\rm d}$, and so
we assume that the optical depth of the disk varies as $1/\mathrm{cos}\theta_{\rm d}$
(see \citet{vanSoelen2012} for a more accurate treatment of the inclined disk case).
This simplified emission model with the power-law prescriptions suffices
to compute the disk spectrum in our model.

The disk parameters for J0632 are not known, and thus we adopt their typical ranges  \citep[][]{Carciofi2006, Klement2017}
to match the observed data of J0632 (Fig.~\ref{fig:fig3}).
The observed spectrum of an isothermal ($s_d=0$) hydrogen disk
extending to 60$R_*$
with a central density $\rho_0=10^{-11}\rm \ g\ cm^{-3}$, disk temperature $T_d=0.7T_*$,
$n_d=2.45$, and $\beta_{\rm d}=1.5$ is presented in Figure~\ref{fig:fig3}.
The disk emission amplitude parameters (e.g., $\rho_0$) used
to match the observed IR-to-optical measurements depend on the assumed disk-viewing angle ($i_{\rm d}$)
between the LoS and the surface normal vector of the disk.
For the measured IR-to-optical flux densities (Fig.~\ref{fig:fig3}),
the `intrinsic' disk emission would be inferred to be stronger for a larger $i_{\rm d}$;
then the disk will provide
the preshock and IBS with a larger amount of ICS seeds. Because it was suggested
that the disk of J0632 is seen nearly edge-on \citep[e.g.,][TAH21]{Aragona2010}, we
assume a large $i_{\rm d}$ of 85$^\circ$ to test a limiting case.

We verified our simple disk emission model by comparing the modeling results
with those of \citet{Carciofi2006} and \citet{vanSoelen2012}.
Firstly, our model for the spectrum of the Be disk in J0632 (Fig.~\ref{fig:fig3})
is similar to that in another TGB PSR~B1259$-$63 \citep[][]{vanSoelen2012}.
Secondly, we found that surface brightness of the disk at high frequencies ($10^{14}$\,Hz)
is significantly higher in the inner region ($r\lapp 2R_*$) than the outer regions, which
is consistent with a previous result \citep[e.g., Fig.~13 of][]{vanSoelen2012}.
Thirdly, emission computed by our model for a moderately inclined disk does not deviate
much from that for a pole-on disk, similar to previous results \citep[][]{Carciofi2006,vanSoelen2012}.
These verify that our simplified disk-emission model captures the main emission features of
Be disks.

\subsection{Synchrotron radiation}\label{sec:sec3_2}

\begin{figure}
\centering
\includegraphics[width=82mm]{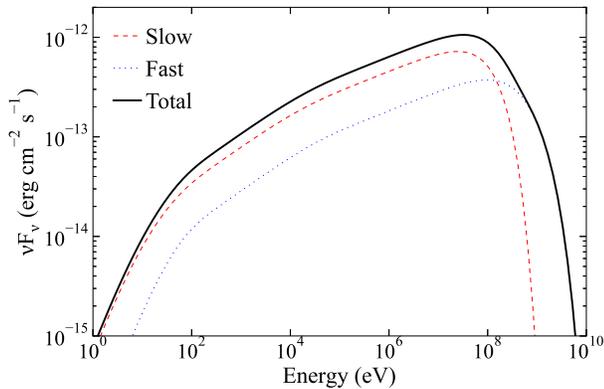}
\figcaption{A phase-averaged synchrotron SED computed for a circular orbit.
The model parameters are detailed in Section \ref{sec:sec3_2}.
Emissions of the slow and fast flows
are plotted in red and blue, respectively, and the summed emission is presented in black.
\label{fig:fig4}}
\end{figure}
	
\begin{figure*}
\centering
\begin{tabular}{ccc}
\hspace{-6mm}
\includegraphics[width=62mm]{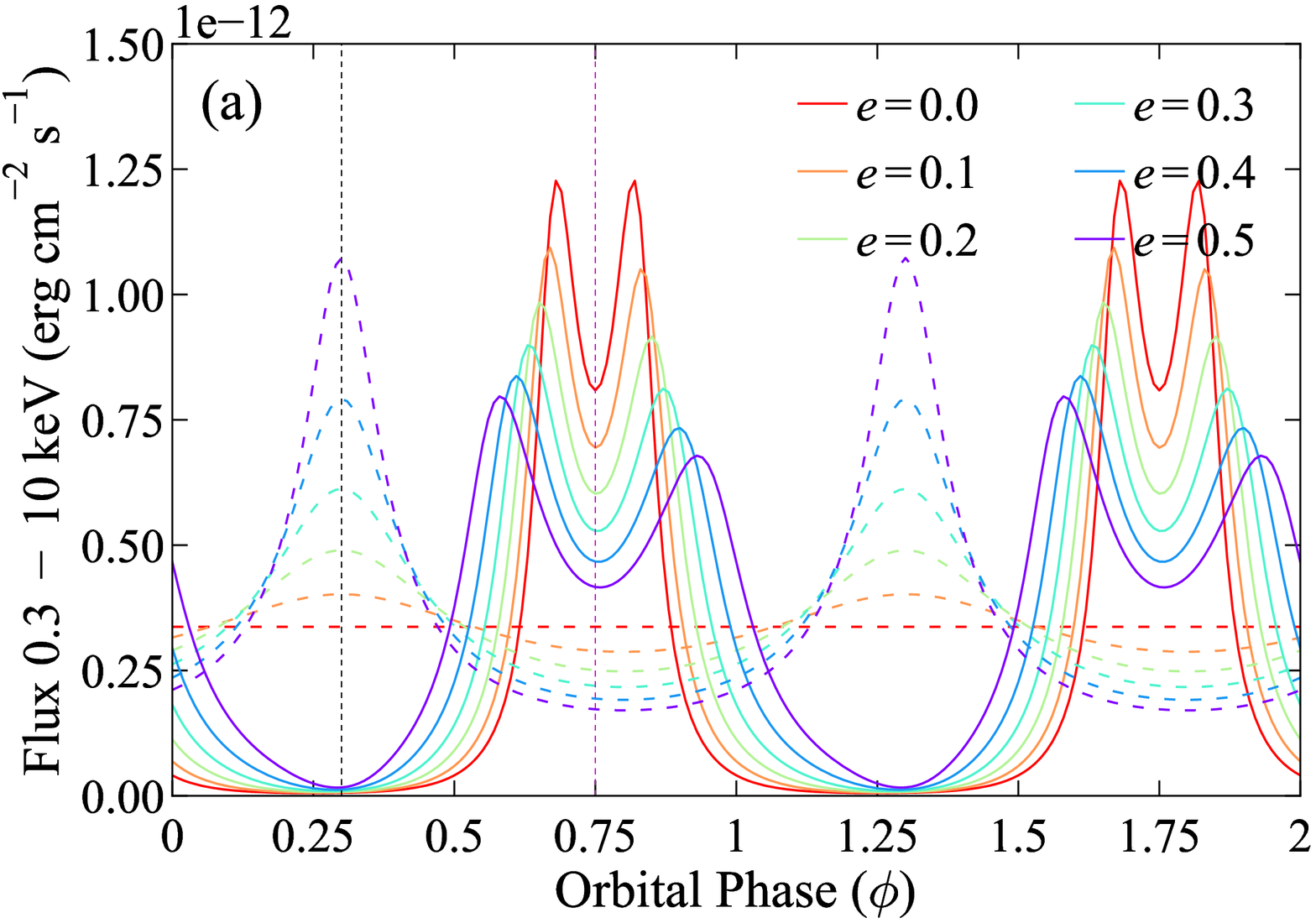} &
\hspace{-6mm}
\includegraphics[width=62mm]{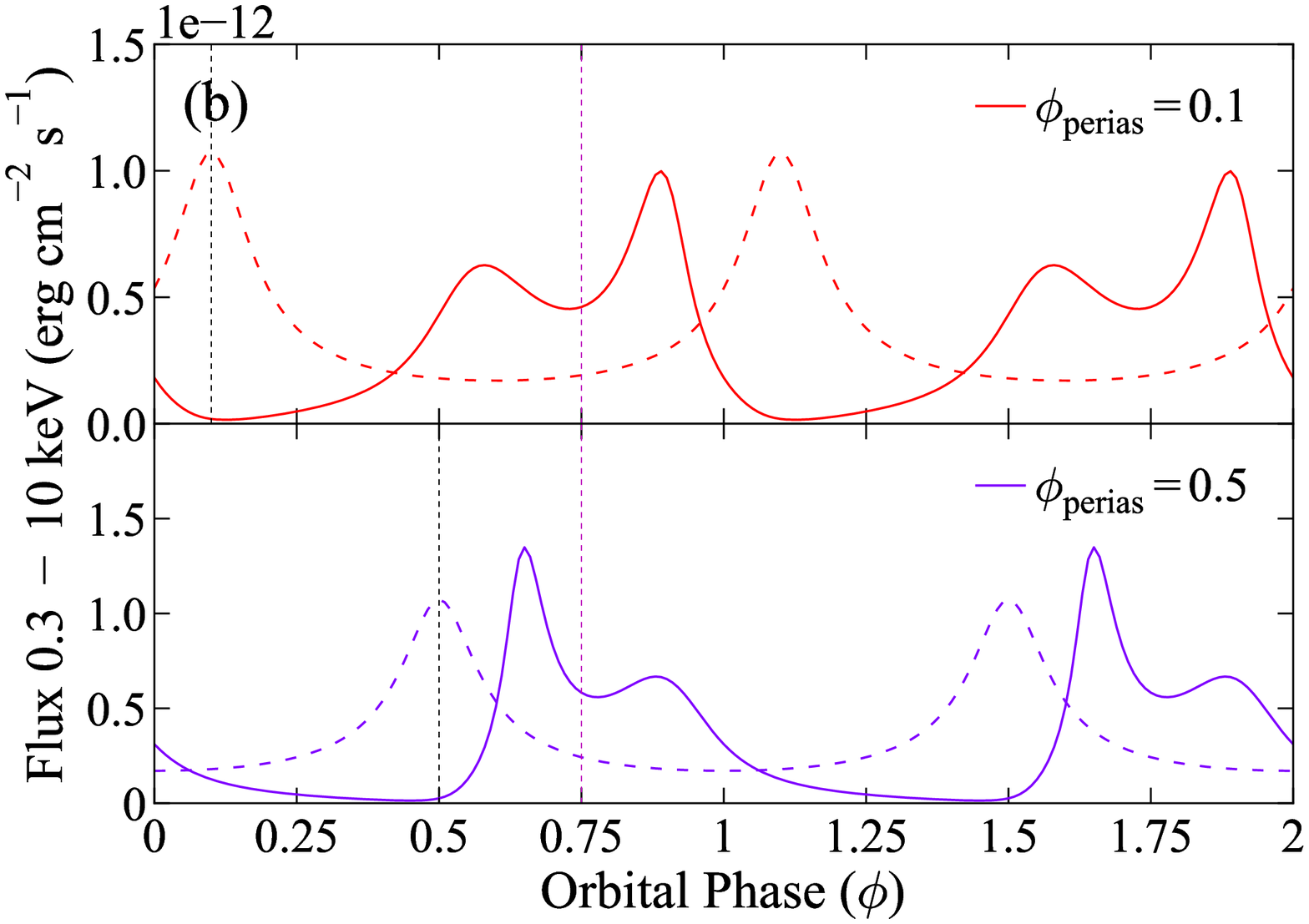} &
\hspace{-6mm}
\includegraphics[width=62mm]{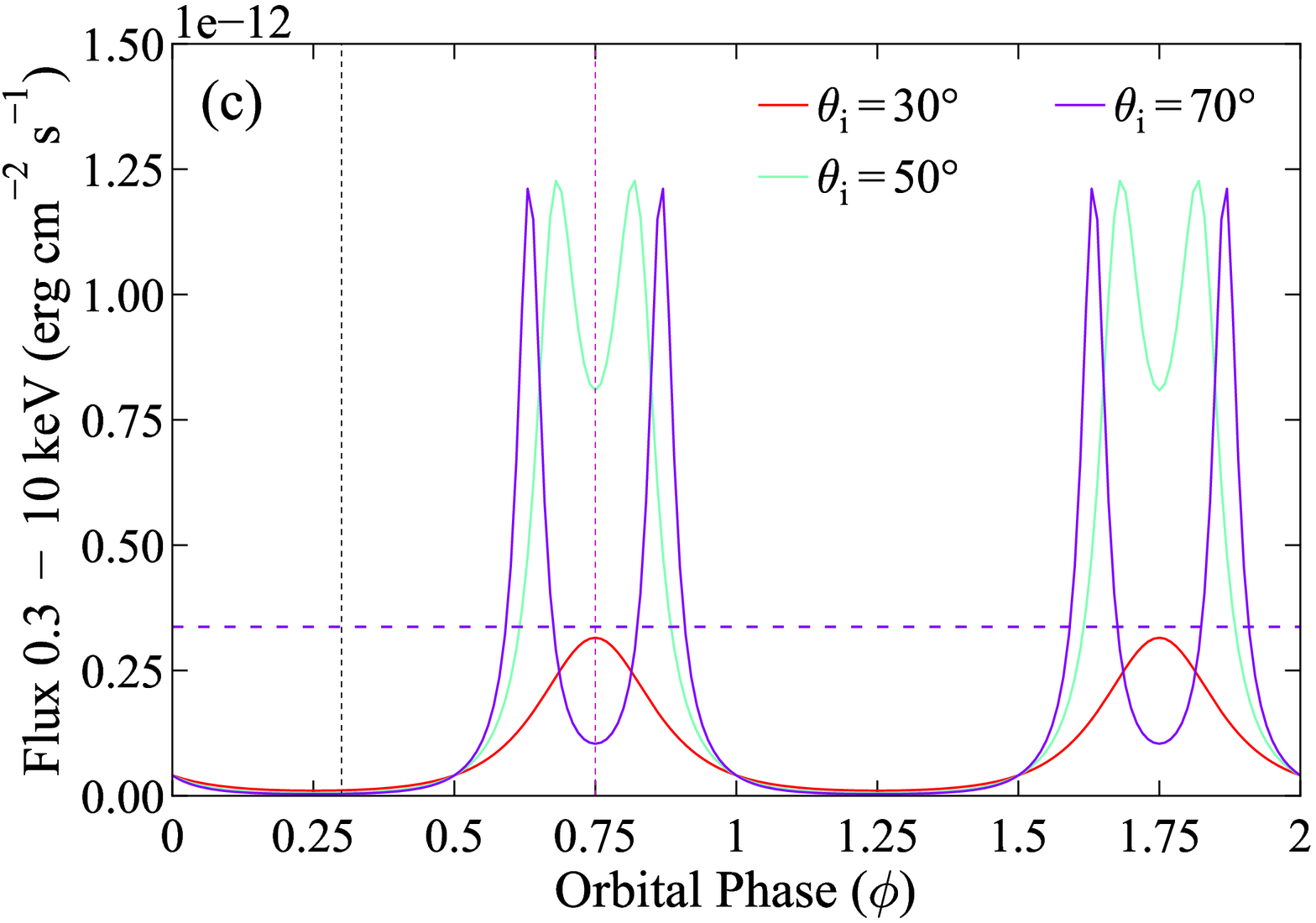} \\
\hspace{-6mm}
\includegraphics[width=62mm]{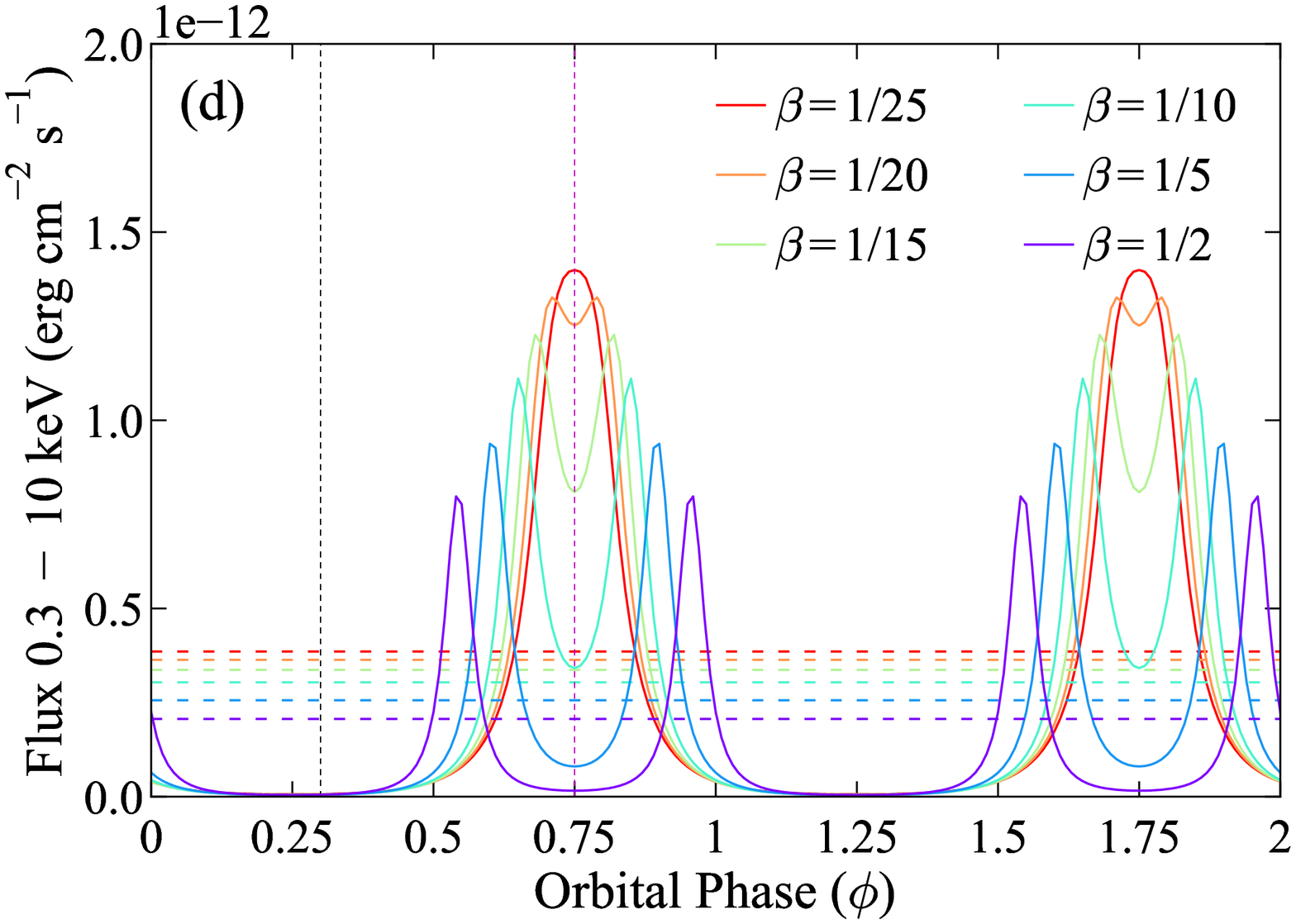} &
\hspace{-6mm}
\includegraphics[width=62mm]{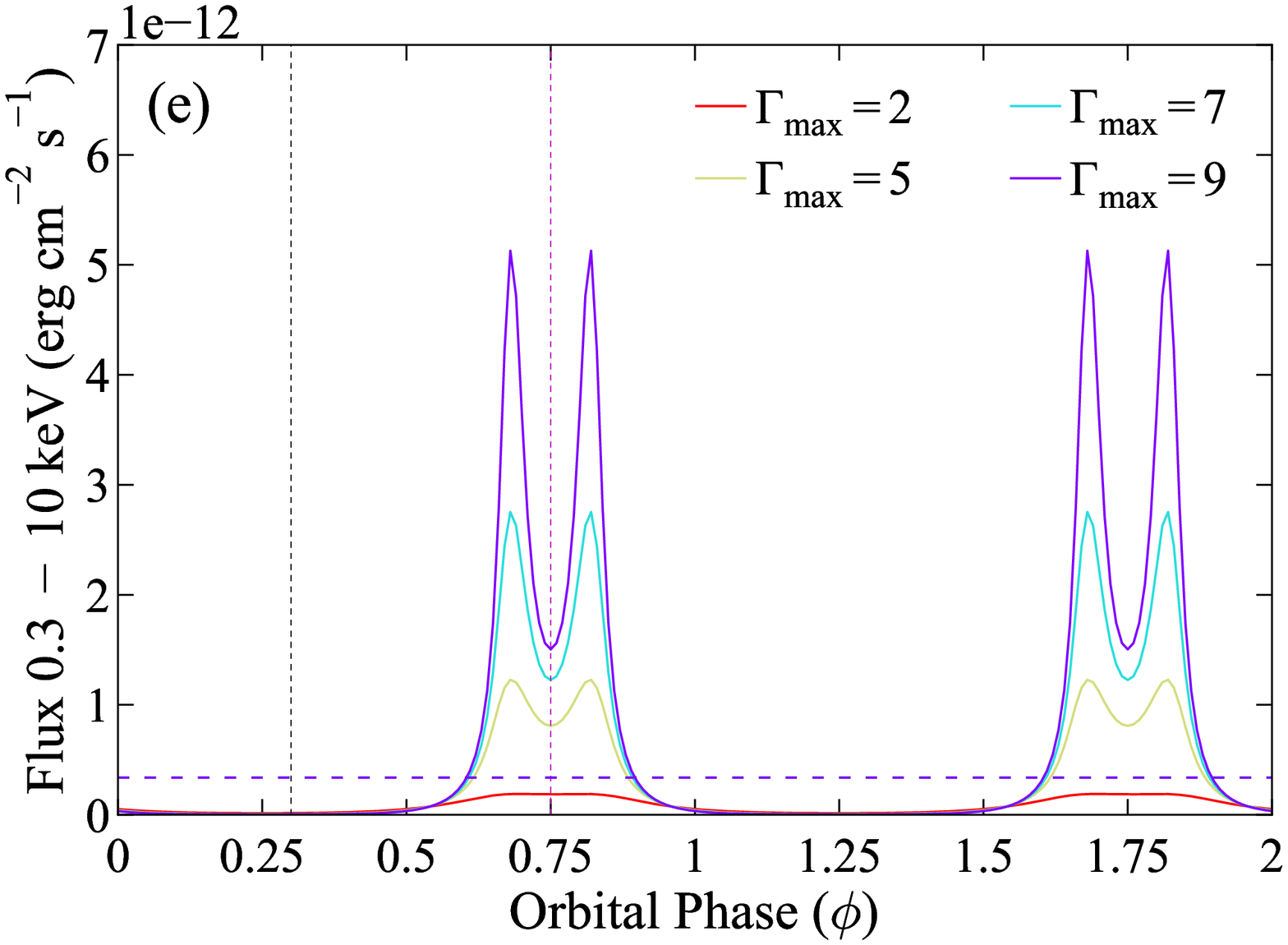} &
\hspace{-6mm}
\includegraphics[width=62mm]{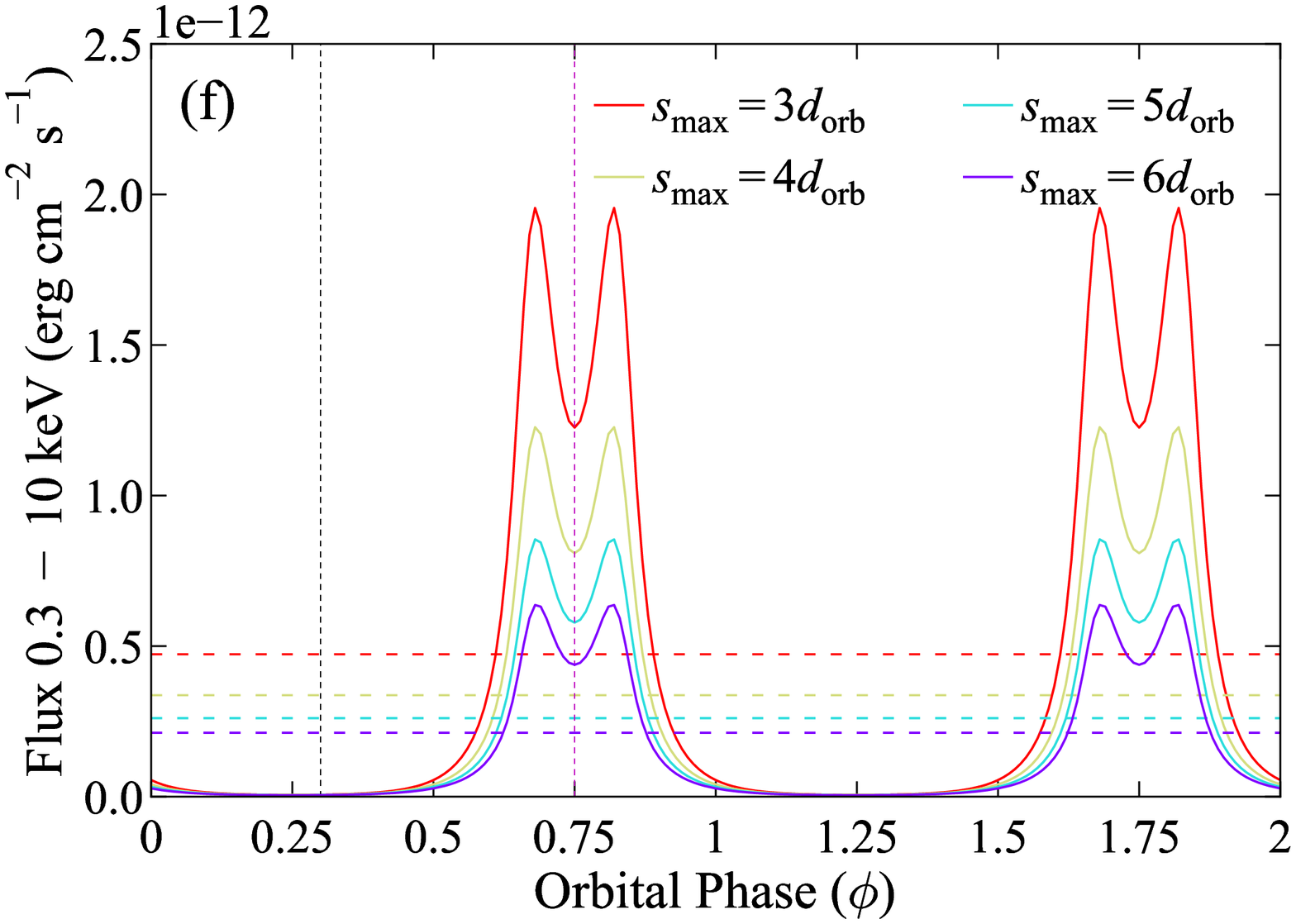} \\
\end{tabular}
\figcaption{SY LC's dependencies on the IBS and orbital parameters.
The baseline parameters are the same as those in Figure~\ref{fig:fig4}, but some of them are varied here.
The IFC phase is 0.75 in all the panels (magenta vertical lines),
and the periastron phase (black vertical lines) is 0.3 except for panel (b).
LCs of the slow and the fast flows are presented in dashed and solid lines, respectively,
and different color denotes different values of the parameter varied for an investigation
in each panel (red to purple from a small to large value).
(a) $e$ dependence of the SY LC. Note that $B_0$ is defined to be 0.28\,G for the baseline circular orbit. For eccentric orbits,
we varied $B_0$ (defined at the shock nose at IFC) by scaling with $r_{\rm p}$ (e.g., Eq.~\ref{eq:Bfield}).
(b) LCs produced using $\phi_{\rm perias}=0.1$ (top) and $\phi_{\rm perias}=0.5$ (bottom) for an eccentric orbit with $e$=0.5.
(c) LCs seen by observers at
different inclinations. (d--f) LCs constructed for various wind momentum flux ratios $\beta$ (d),
$\Gamma_{\rm max}$ (e), and $s_{\rm max}$ (f).
\label{fig:fig5}}
\end{figure*}
	
	With particle distributions and $B$ in IBS being prescribed (Sections~\ref{sec:sec2_3}--\ref{sec:sec_2_5}),
a SY SED is computed using the formulas given in Appendix~\ref{sec:appendix3} \citep[see][for more detail]{Finke2008}.
We construct an IBS surface at each of 100 orbital phases and divide it into
$21\times 361$ (axial and azimuthal) emission zones. The particle distribution is computed in
300 energy ($\gamma_e$) bins and the resulting SED is computed in 200 frequency bins.
These binnings were chosen to
achieve sufficient accuracy to fit the observation data while maintaining reasonable computation time.
As an example, we consider a circular orbit with
orbit radius $a=3.5\times 10^{13}$\,cm, periastron at $\phi_{\rm perias}=0.3$
and the inferior conjunction (IFC) of the pulsar at $\phi_{\rm IFC}=0.75$.
For the IBS, we used $\beta=0.07$, $p_1=2.3$, $p_2=2.6$, $\gamma_{e,\rm min}=1.5\times 10^5$,
$\gamma_{\rm b}=5\times 10^6$, $s_{\rm max}=4d_{\rm orb}$, $\xi=0.1$, $B_0$=0.28\,G, $\Gamma_{\rm max}=5$,
and $\theta_i=50^\circ$. Using these baseline values,
we compute parameters in the IBS flow (see Fig.~\ref{fig:fig2}), e.g., $N_e$ (Eq.~\ref{eq:IBSparticles}),
$\Gamma$ and $\delta_{\rm D}$ (Eqs.~\ref{eq:bulkG} and \ref{eq:Doppler}),
$B$ (Eq.~\ref{eq:Bfield}). Then the SY emission SED (Eq.~\ref{eq:sysed}) in each zone is computed
neglecting the SY emission of the `cold' preshock electrons.

    Figure~\ref{fig:fig4} shows an orbitally-averaged SED.
The computed SED matches the expected emission features at energies $10^2$\,eV, $10^4$\,eV and $10^8$\,eV, 
which correspond to  $\gamma_{e,\rm min}$, $\gamma_{\rm b}$, and $\gamma_{e,\rm max}$, respectively.
Even though the fast flow is assumed to be only a small fraction ($\xi=0.1$) of the slow flow,
the fast-flow emission is highly boosted near the IFC
and is noticeably strong in the phase-averaged SED.
The shape of the SY SEDs of the two flows are similar, but that of the fast flow
is shifted to higher energies (blue in Fig.~\ref{fig:fig4}) because of Doppler
beaming by the bulk motion (Eq.~\ref{eq:Doppler}).

	The SY LC of IBS flows is influenced by the orbital and the IBS parameters.
Orbital variation of the `slow' flow is induced
by varying $r_{\rm p}$ in an eccentric orbit as
$B\propto 1/r_{\rm p}\propto 1/d_{\rm orb}$ (Eqs.~\ref{eq:Bfield} and \ref{eq:shocknose}).
Hence, LCs of the slow flow (dashed lines in Fig.~\ref{fig:fig5}a)
have a peak at periastron $\phi_{\rm perias}=0.3$.
Because SY flux ($F_{\rm SY}$) is proportional to $B^{(p_1 + 1)/2}$ \citep[][]{Dermer1995,An2017},
we anticipate the orbital variation being $\propto 1/d_{\rm orb}^{(p_1 + 1)/2}$. For eccentricity $e$,
the min-max ratio in the LC of the slow flow (e.g., dashed lines in Fig.~\ref{fig:fig5} a)
would be $\propto \left ( \frac{1 + e}{1 - e} \right )^{(p_1 + 1)/2}$ (e.g., Eq.~\ref{eq:dorb}).

    Emission of the `fast' flow arises mostly from the tail ($s\approx s_{\rm max}$)
of the shock where $\Gamma$ and the particle density are largest (Eqs.~\ref{eq:bulkG} and \ref{eq:IBSparticles}).
Because IBS particles flow along
a cone-shape surface (blue in Fig.~\ref{fig:fig2}) and the emission of
the fast flow is highly beamed in the flow direction,
the sky emission pattern of the fast flow will be  ring-like in the shock tail direction \citep[e.g.,][]{Romani2016}.
Due to different viewing angles of the shock tail, the Doppler factor will vary with the orbital phase. Since $F_{\rm SY}\propto \delta_{\rm D}^{(5 + p_1)/2}$ \citep[e.g.,][]{Dermer1995,An2017},
 the observed flux of the fast flow will be largest
near the IFC
where the flow direction aligns well with the LoS (e.g., Fig.~\ref{fig:fig2} left).
Note that $B$ ($\propto 1/d_{\rm orb}$) also varies with $\phi$ and affects the amplitude of the variation,
but this effect is small compared to the $\delta_{\rm D}$ effect for the fast-flow emission.
Hence, its peak occurs at the IFC.

	Figure~\ref{fig:fig5} shows the SY LC's dependencies on the orbital and IBS parameters.
Panel (a) shows an effect of the eccentricity ($e$). For a circular orbit ($e=0$),
a modulation of the slow-flow emission is very weak
(i.e., weak beaming due to low $v_{\rm flow}$); it was ignored in this example for clarity.
The fast-flow emission modulation is large even in the $e=0$ case.
In an eccentric orbit, the phase separation between periastron and the IFC also affects the shape of the SY LC
(e.g., Fig.~\ref{fig:fig5} b with an assumed $e$=0.5).
The double peaks of the fast-flow emission (solid line) may have
substantially different amplitudes and widths; one closer to the periastron
is higher and sharper because of larger $B$ and rapid orbital motion of the pulsar.

	The effects of $\theta_i$, $\beta$, $\Gamma_{\rm max}$, and $s_{\rm max}$
are presented in Figure~\ref{fig:fig5} (c)--(f), respectively.
For a given IBS opening angle ($\theta_{\rm cone}$),
$\theta_i$ determines the viewing angle ($\theta_{\rm V}$) and thereby $\delta_{\rm D}$ (Eq.~\ref{eq:Doppler}) for the emission.
For $\theta_i=0$, circular orbits do not cause a  modulation of 
the slow or the fast flow. As $\theta_i$ grows,
the LoS becomes closer to the shock tangent (i.e., emission ring) near the IFC, and the modulation induced by orbitally varying $\theta_{\rm V}$ should be observed.
For a sufficiently large $\theta_i$ (i.e., $\theta_i>\pi/2-\theta_{\rm cone}$; Fig.~\ref{fig:fig2}),
the LoS crosses the emission ring twice near IFC,
and the fast-flow emission bump in the LC splits into two peaks (red vs. purple in Fig.~\ref{fig:fig5} c);
the separation between them (LoS crossings of the emission ring) increases with $\theta_i$ and
is $2\theta_{\rm cone}$ for $\theta_i=90^\circ$.

\begin{figure}
\centering
\includegraphics[width=70mm]{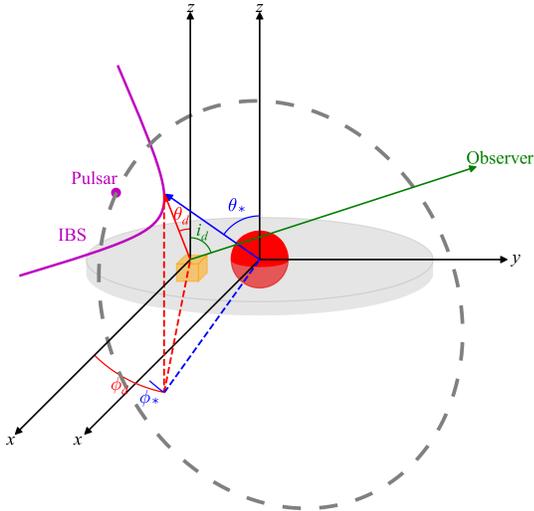} \\
\figcaption{Geometry for ICS scattering off of the stellar and disk photons.
The red circle is the stellar companion with an equatorial disk, shown as a gray disk.
The pulsar is denoted by a purple dot, and the purple parabolic
curve around it depicts the IBS. Both the pulsar and stellar companion are assumed to be point sources, while 
the orange cube represents a small volume element of the disk.
The orbit of the pulsar is displayed in a gray dashed line.
In this coordinate, the observer is in the y direction at an inclination angle of $i_{\rm d}$ (disk).
The incident angles of the stellar and disk seed photons into the emission zone on the IBS are shown as 
$(\theta_*,\phi_*)$ and $(\theta_{\rm d},\phi_{\rm d})$, respectively.
Note that these angles vary over both the IBS cone and the stellar and disk surfaces.
In this work, we assume that the incident angles into an emission zone of the IBS
are the same over the stellar and disk surfaces,
but the change of the angles into different zones of the IBS cone is taken into account.
See text for more detail.
\label{fig:figdisk}}
\end{figure}

Effects of $\beta$ (Fig.~\ref{fig:fig5} d) are similar to those of $\theta_i$
since $\beta$ determines $\theta_{\rm cone}$ (Eq.~\ref{eq:coneangle})
and thus $\theta_{\rm V}$. However, separation of the two peaks in the fast-flow LC
depends more sensitively on $\beta$ than $\theta_i$. This is an obvious
geometrical effect of the former ($\beta$) directly
changing the emission-ring size $\theta_{\rm cone}$.
Note also that $r_{\rm p}$ (see Fig.~\ref{fig:fig2}) is determined by $\beta$
(e.g., Eq.~\ref{eq:shocknose}); for a smaller $\beta$,
$r_{\rm p}$ is smaller and thus $B$ ($\propto 1/r_{\rm p}$) in the IBS is stronger, increasing the
SY flux of the slow flow as well (dashed lines in Fig.~\ref{fig:fig5} d).
The bulk Lorentz factor of the fast flow near the shock tail (i.e., $\Gamma_{\rm max}$)
controls the width of each peak (i.e., $1/\Gamma$ beaming)
as well as emission strength of the fast flow (Fig.~\ref{fig:fig5} e).

\begin{figure*}
\centering
\includegraphics[width=180mm]{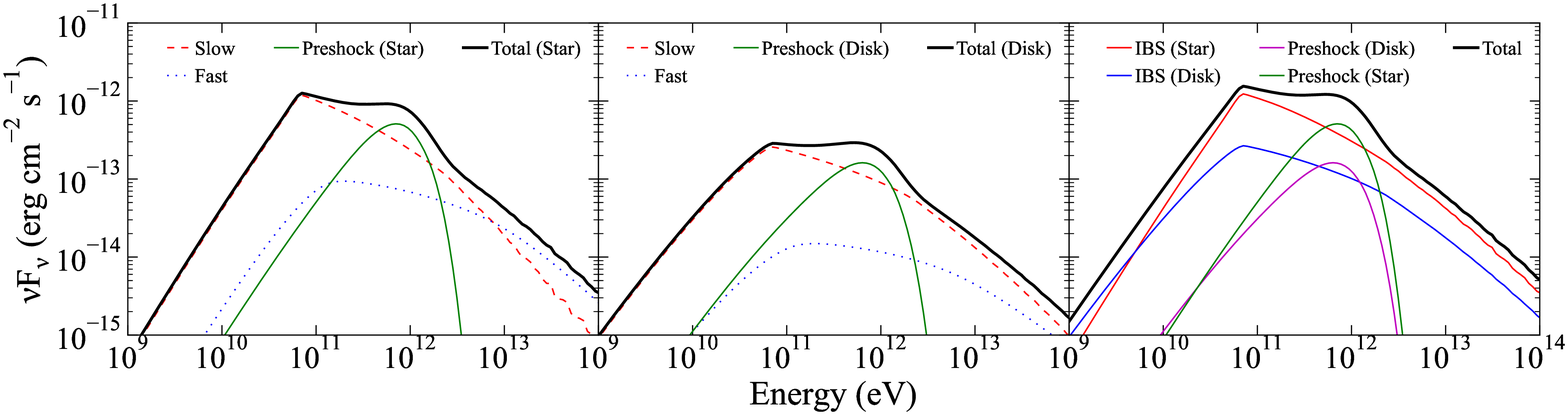}
\figcaption{Models for VHE SEDs produced by
the external Compton (EC) process off of the stellar blackbody emission (left), 
disk emission (middle), and their sum (right).
In the left and middle panels, EC emissions of electrons in the slow and the fast flows of IBS,
and in the preshock flow are shown in red dashed, blue dotted, and green solid lines,
respectively. In the right panel, IBS EC emissions off of the stellar and disk photons are
presented in red and blue solid lines, respectively, and preshock emissions off of the
stellar and disk photons are shown in green and purple, respectively.
\label{fig:fig6}}
\end{figure*}

	The emission strengths of the slow- and the fast-flow particles change
depending on the length of IBS ($s_{\rm max}$) even if the particle number does not change
(Fig.~\ref{fig:fig5} f). This is because of the decrease in $B$
at large distances from the pulsar ($r_{\rm p}$); the total IBS emission is reduced 
with increasing $s_{\rm max}$. This effect is more pronounced for the fast-flow emission
because it mostly arises from the shock tail. Thus the peak-amplitude ratio of the fast-
to the slow-flow emissions moderately decreases with increasing $s_{\rm max}$ as is 
noticeable in Fig.~\ref{fig:fig5} (f).

\subsection{ICS Emission}\label{sec:sec3_3}
	VHE emission can be produced by ICS processes: the synchrotron self-Compton (SSC) process in the IBS and
the external Compton (EC) process of the preshock and IBS particles. 
However, the electron density in TGBs is too low to produce significant SSC emission flux in the IBS. Hence, the SSC process is often ignored for TGBs and we therefore only consider the EC process for VHE emission.
We assume a head-on collision for the EC scattering which is appropriate
for ultra-relativistic particles ($\gamma_e\gg1$) in the preshock and IBS,
and calculate their emission SEDs using formulas given in Appendix~\ref{sec:appendix3}
\citep[see][for more detail]{Dermer2009}.

\begin{figure*}
\centering
\begin{tabular}{ccc}
\hspace{-6mm}
\includegraphics[width=62mm]{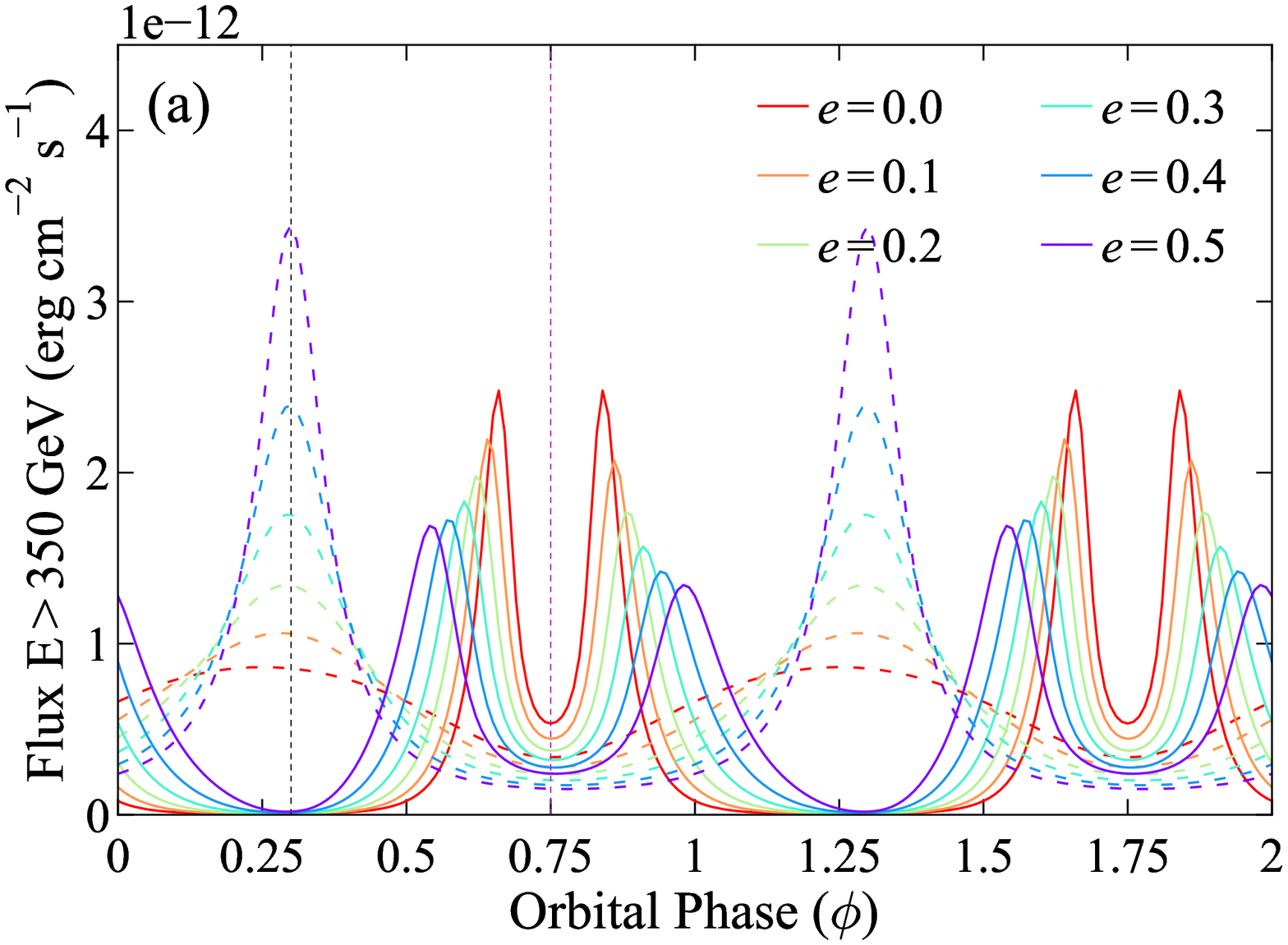} &
\hspace{-6mm}
\includegraphics[width=62mm]{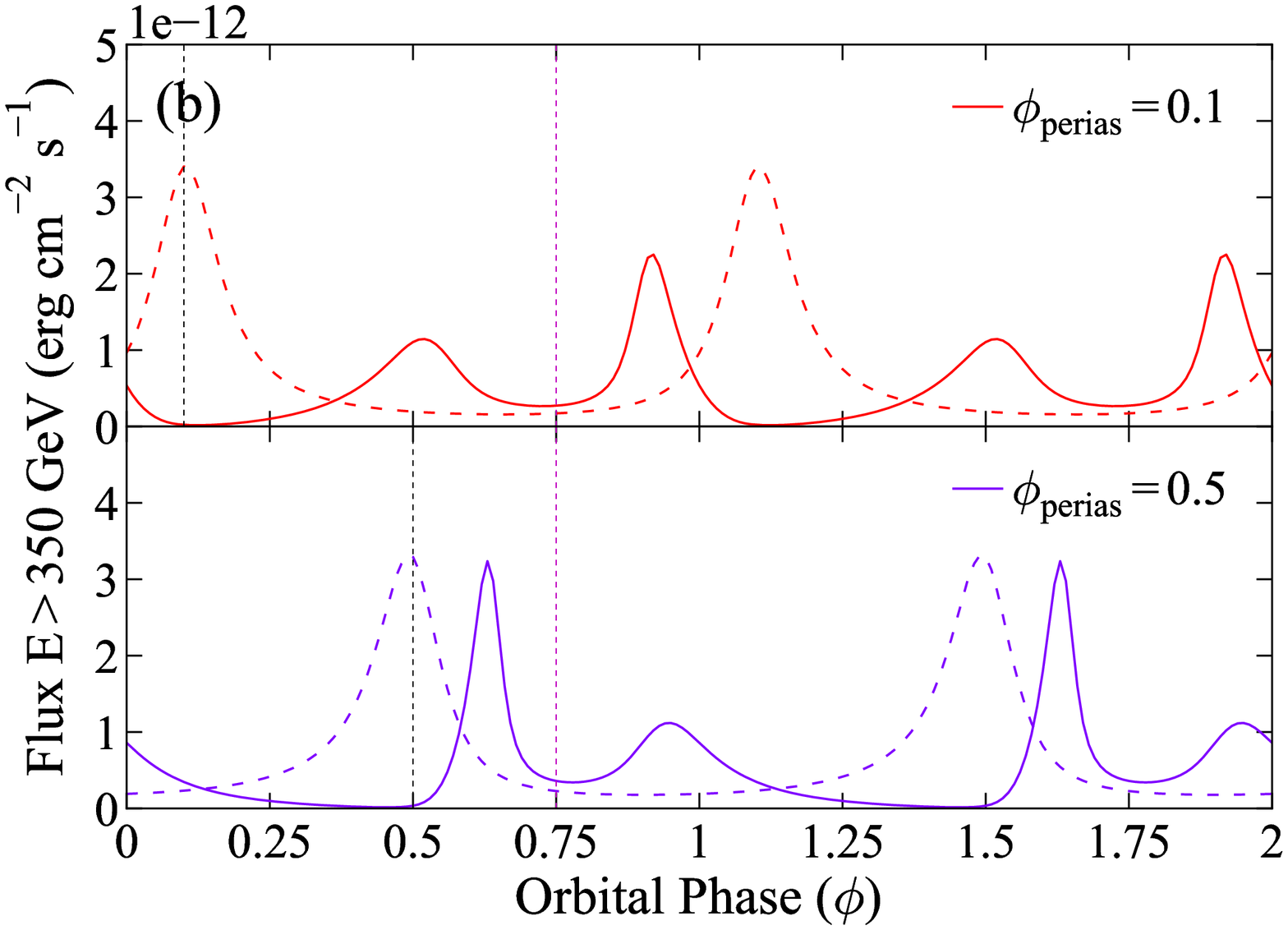} &
\hspace{-6mm}
\includegraphics[width=62mm]{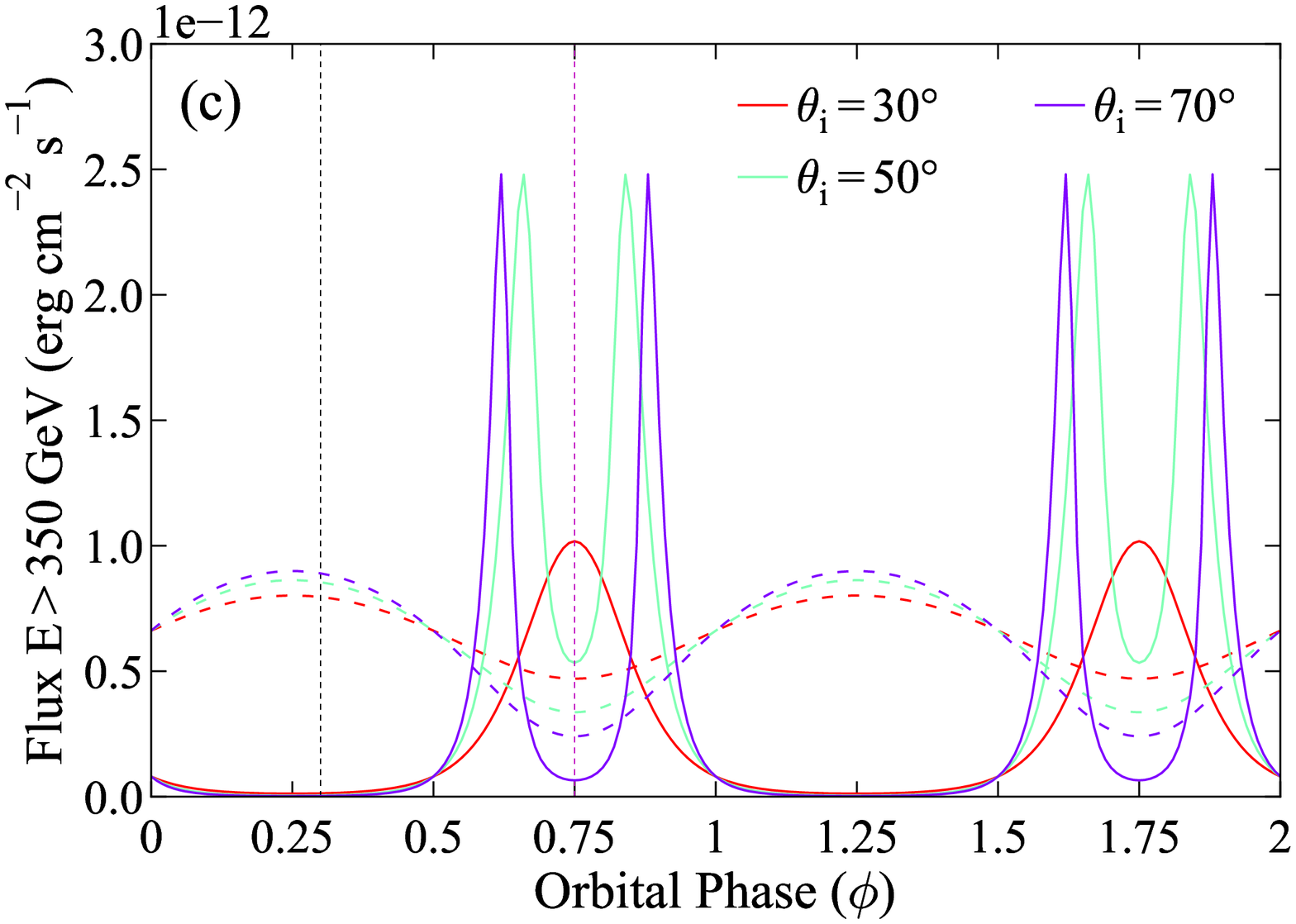} \\
\hspace{-6mm}
\includegraphics[width=62mm]{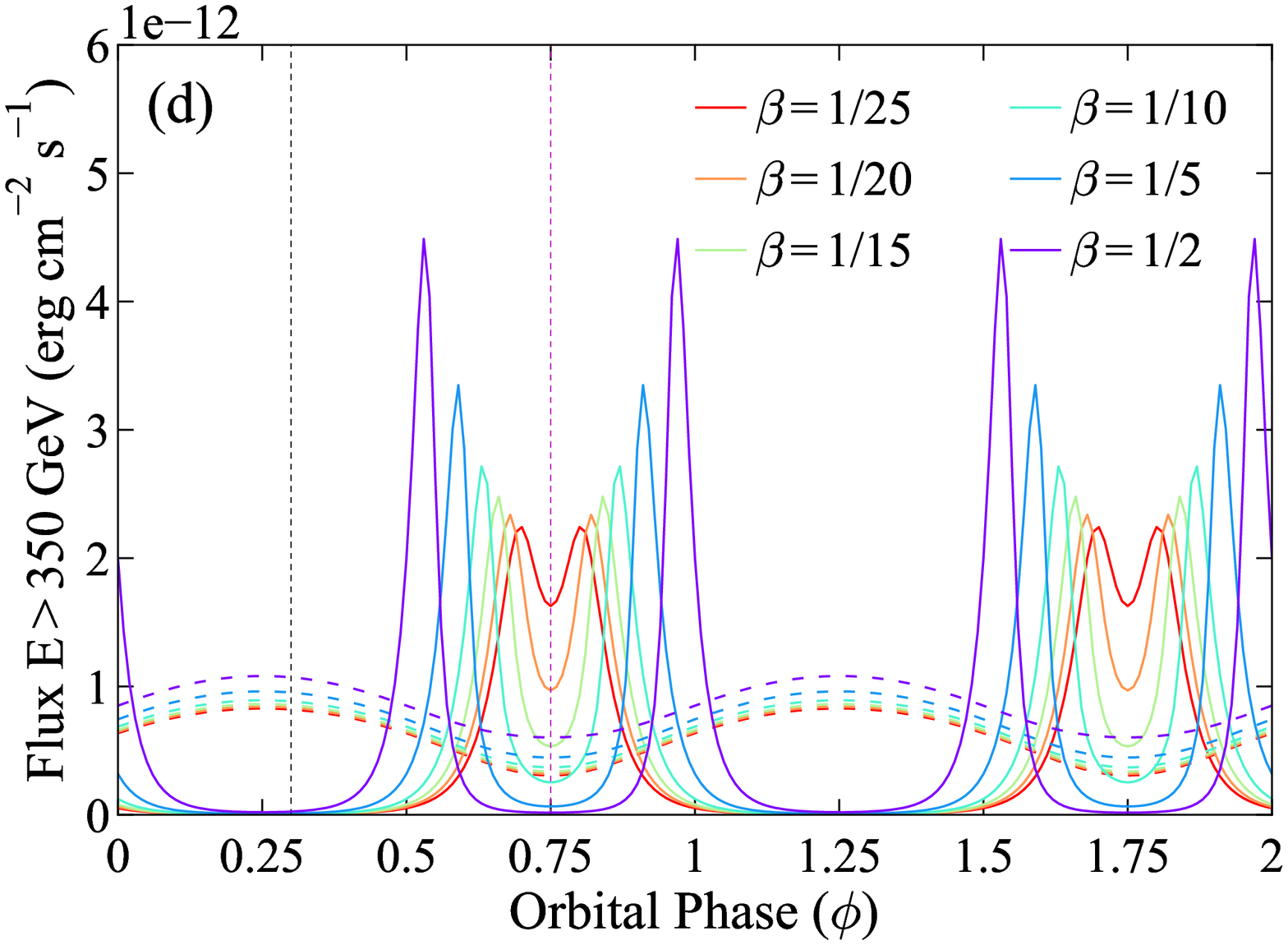} &
\hspace{-6mm}
\includegraphics[width=62mm]{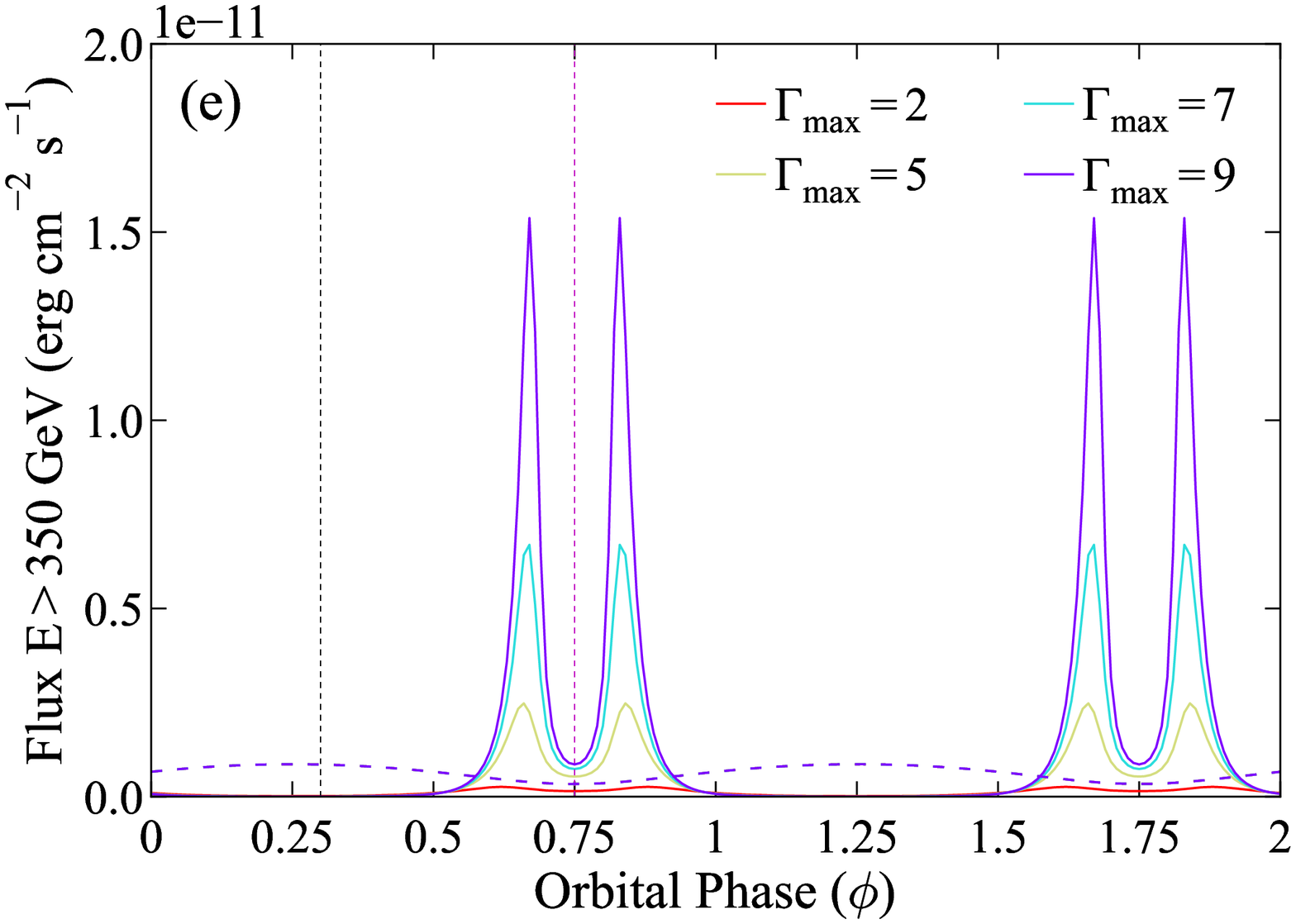} &
\hspace{-6mm}
\includegraphics[width=62mm]{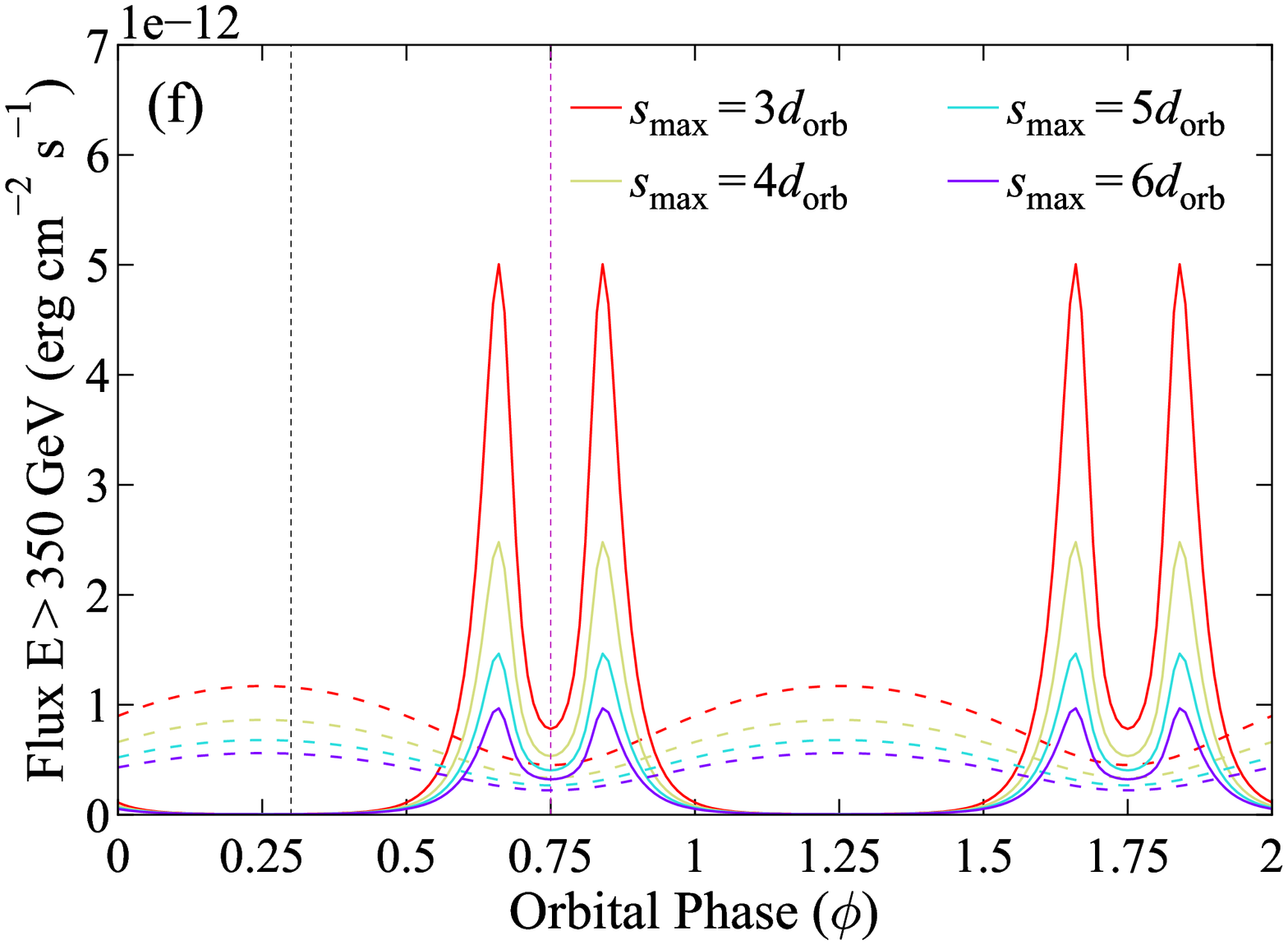} \\
\end{tabular}
\figcaption{EC LCs constructed with the same parameters as
those in the corresponding panels of Figure~\ref{fig:fig5}.
LCs of the slow and the fast flows are presented in dashed and solid lines, respectively,
and different color denotes different values of the parameter varied
in each panel (red to purple from a small to a large value).
Vertical lines mark periastron (0.3 except for panel b; black) and IFC (0.75; magenta) phases.
(a) LC's dependence on $e$. (b) LCs produced using
$\phi_{\rm perias}=0.1$ (top) and $\phi_{\rm perias}=0.5$ (bottom). (c) LCs observed at
different inclinations. (d--f) LCs constructed for various $\beta$ (d),
$\Gamma_{\rm max}$ (e), and $s_{\rm max}$ (f).
\label{fig:fig7}}
\end{figure*}
	
	For the stellar blackbody seeds, we assume that the star is a point source.  
Therefore, both the incidence azimuth and polar angles $\phi_*$ and
$\theta_*$ ($\mu_*\equiv \mathrm{cos}\theta_*$) of the seed photons
(see Eq.~\ref{eq:ecsed}) into a scattering emission zone,
as well as the distance ($r_{\rm s}$) from the stellar surface to the scattering zone in the IBS or preshock,
are the same over the surface of the companion (see Fig.~\ref{fig:figdisk}).
This is a reasonable approximation for the companion of J0632 since
$R_* \ll d_{\rm orb}$. In contrast, the size of the disk can be comparable to the orbit, and thus
the incident azimuth and polar angles $\phi_{\rm d}$ and $\theta_{\rm d}$
($\mu_{\rm d}\equiv\mathrm{cos}\theta_{\rm d}$)
of the disk seeds into the scattering zone vary over the disk surface (Fig.~\ref{fig:figdisk}), 
as does the
distance from a surface element of the disk to the scattering zone ($r_{\rm d}$).
Additionally, the disk emission is anisotropic
and varies radially (Section~\ref{sec:sec3_1}). 
These variations require computationally demanding calculations, since the 
EC computation is carried out
in each of $21\times 361$ IBS zones, 500 preshock zones, and 100 orbital phases, in addition to 300 energy ($\gamma_e$) bins.
Hence, we simplified EC computations by assuming the disk is a point source with 
$\phi_{\rm d}=\phi_*$, $\mu_{\rm d}=\mu_*$, and $r_{\rm d}=r_{\rm s}$ over the
disk surface.
Note, however, that the EC scattering angles in and distance from the disk to a scattering zone
vary over the IBS surface because the location and the flow direction
of each zone change (Figs.~\ref{fig:fig2} and \ref{fig:figdisk}).
We verified that EC spectra computed with this  assumption were
not significantly different from those obtained with a full integration over
the disk surface at a few phases. Further note that the disk emission is
weaker than the companion's blackbody emission (see Fig.~\ref{fig:fig3}), and therefore
contribution of the disk seeds to the EC SED is small (Fig.~\ref{fig:fig6}).

	With the aforementioned assumptions, we compute the EC emission.
In each emission zone, we compute the incidence angles and spectral energy densities ($u_*$)
of the stellar and disk emissions, calculate the EC emission (Eq.~\ref{eq:ecsed}) in the zone, and
integrate over the IBS surface to generate the EC SED at a phase.
Figure~\ref{fig:fig6} shows orbitally-averaged EC SEDs constructed using the same parameters
as those used for Figure~\ref{fig:fig4}.
Note that the EC SEDs also reflect the distributions of emitting particles and seed photons;
broader distributions result in a more extended EC SED. 
Hence, the EC SED for the disk photons (Fig.~\ref{fig:fig6} middle) is slightly broader
than that for the stellar photons (Fig.~\ref{fig:fig6} left).
Notice that any sharp features in the disk-seed spectrum (Fig.~\ref{fig:fig3})
are not apparent in the EC SED as they are blurred by the broad electron distributions.

	The EC emission of IBS particles varies orbitally due to changes
in the seed photon density, the scattering angle $\psi$ (the angle between incoming and scattered photons),
and Doppler beaming of the emission.
The seed photon density varies as $\propto 1/r_{\rm s}^2$ over the orbit, and so the orbital
modulation of the EC emission from the slow flow is similar to
that of the SY emission (dashed curves in Fig.~\ref{fig:fig5}).
As in the SY case, the EC emission of the fast flow is strongest
at IFC where the flow direction is well aligned with LoS thereby leading to the strongest Doppler beaming.
However, the EC LC varies in a more complex way because of changes in the scattering
geometry (e.g., scattering angle $\psi$) over the orbit which is most favorable
near superior conjunction of the pulsar (SUPC; $\psi\approx \pi$).

	Dependencies of the EC LCs on the orbital and IBS parameters are similar to
those of the SY LC. We use the same parameters as were used for Figure~\ref{fig:fig5}
and compute EC LCs. The results are displayed in Figure~\ref{fig:fig7}.
In the EC case, the slow flow emission exhibits strong modulation even in
circular orbits due to variation of the scattering geometry (red dashed line in Fig.~\ref{fig:fig7} a).
In addition, the seed photon density for ICS in eccentric orbits is highest at
periastron due to small $r_{\rm s}$, and thus
emission of the slow flow is further enhanced at that phase (dashed lines in Fig.~\ref{fig:fig7}a).
The peaks in the fast-flow EC LCs appear sharper than the corresponding
peaks in the SY LCs (Fig.~\ref{fig:fig5}). This is due to the strong
dependence of EC emission on $\delta_D$ \citep[$\delta_D^{3+p_1}$;][]{Dermer1995}
with small contribution of the $\psi$ effect.
$\beta$ dependence of the strengths of the SY and the EC emissions is opposite
to each other (Fig.~\ref{fig:fig5}d vs. Fig.~\ref{fig:fig7}d)
as a smaller $\beta$ pushes the IBS closer to the pulsar (higher $B$) but
farther from the companion (lower $u_*$).
$s_{\rm max}$ dependence of the EC emission (Fig.~\ref{fig:fig7} f) is similar to that of
the SY emission, 
but the ratio of the fast flow to slow flow fluxes at the IFC drops with 
$s_{\rm max}$ faster in the EC regime than in the SY regime.
This is produced by a combination of $r_{\rm s}$ and $\psi$ changes.
Consequently, the IBS size ($s_{\rm max}$) can be estimated by comparing slow-to-fast-emission
ratios of the SY and the EC emissions.

\subsection{$\gamma$-$\gamma$ absorption}\label{sec:sec3_4}
\begin{figure}
\centering
\includegraphics[width=85mm]{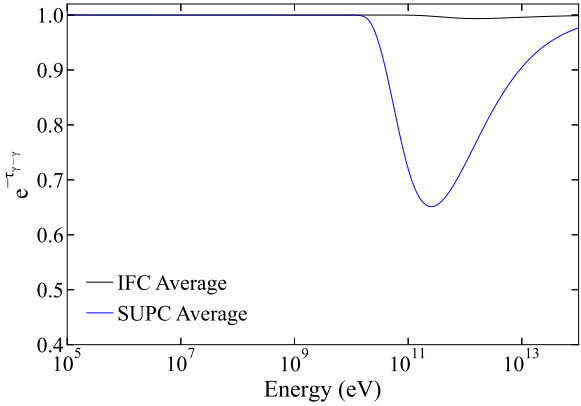} \\
\figcaption{$\gamma$-$\gamma$ absorption by the stellar+disk emission
at IFC (black) and SUPC (blue). Average absorption for emission
over the entire IBS is estimated by comparing the absorbed spectrum to the emitted one.
\label{fig:fig8}}
\end{figure}

	VHE emission is absorbed by soft photons
(stellar blackbody and disk emission) through
the $\gamma$-$\gamma$ pair production process. In each of the IBS and preshock emission zones,
we compute the $\gamma$-$\gamma$ optical depth $\tau_{\gamma\gamma}$
along the LoS using the scattering cross section given in \citet{Gould1967}.
The VHE emission in each zone (Section~\ref{sec:sec3_3}) is then reduced by a factor of $e^{-\tau_{\gamma\gamma}}$
appropriate for the zone. If the orbit is tight, a large companion may block part of IBS or preshock
as was seen in pulsar binaries \citep[e.g.,][]{Corbet2022};
this is not a concern for J0632.

	Examples of $\gamma$-$\gamma$ absorption by the blackbody
and disk emission (in Figure~\ref{fig:fig3}) for parameters appropriate for
J0632 (e.g., Table~\ref{ta:ta1})
are presented in Figure~\ref{fig:fig8}.
As expected, the maximum absorption occurs at SUPC for gamma-ray
photons with $E\approx \frac{(m_e c)^2}{h\nu_{\rm seed}} \approx 10^{11}$\,eV \citep[][]{Gould1967}.
Note that the emission zones are spread over the extended IBS and thus the effect of the absorption
is not very large in the spectrum integrated over the IBS even though the absorption
of the emission at the shock nose (nearest to the companion) would be somewhat stronger.
Secondary electrons produced by the $\gamma$-$\gamma$ interaction may be significant
if $\tau_{\gamma\gamma}$ is large \citep[e.g., $\gg1$;][]{Bednarek2013,Dubus2013} but we do not consider
them in this work since $\tau_{\gamma\gamma}$ in J0632 seems not to be very large.

\section{Modeling the LCs and SEDs of J0632}\label{sec:sec4}
\subsection{Broadband SED and multi-band LCs of J0632}\label{sec:sec4_1}
	We compiled broadband data of J0632 from published papers. 
Its X-ray LCs and spectra have been measured by Swift-XRT and NuSTAR (TAH21),
and a $\sim$GeV SED \citep[][]{Li2017} was measured by the Fermi large area telescope \citep[LAT;][]{Atwood2009}.
Note that the Fermi-LAT flux may include emission of a putative pulsar which we do not model, and thus we regard
the $\sim$GeV flux measurements as upper limits.
In the VHE band, VERITAS and H.E.S.S. data presented in \citet{Adams2021} are used to construct
the LCs and SEDs.

\begin{figure}
\centering
\includegraphics[width=80mm]{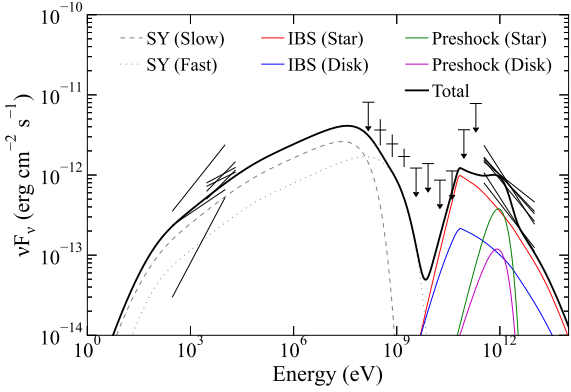}
\figcaption{Broadband SEDs of J0632 and model computations (averaged over the orbit).
The X-ray to VHE data are taken from literature \citep[TAH21;][]{Aliu2014,Li2017,Adams2021}. The black solid line segments
are the best-fit X-ray (Swift and NuSTAR) and
VHE power-law models at
several representative phases \citep[TAH21;][]{Aliu2014,Adams2021},
and black data points are Fermi-LAT measurements \citep[][]{Li2017}.
Model computations are presented by curves;
gray lines are for the SY emissions (dashed for the slow and dotted for the fast flow),
and red and blue solid lines are the EC emissions off of the stellar and disk seeds, respectively.
EC emissions by the preshock particles are denoted
in green and purple for the stellar and the disk seeds, respectively.
The summed model is shown by a black solid curve.
\label{fig:fig9}}
\end{figure}

	For the X-ray SEDs, we plot those measured with Swift in the
0.3--10\,keV band and with NuSTAR
in the 3--20\,keV band. Since the X-ray spectrum of J0632 is variable,
we take three representative power law models (i.e., three orbital phases)
in each of the Swift and the NuSTAR bands \citep[Fig.~\ref{fig:fig9};][TAH21]{Aliu2014}.
For the VHE SEDs, we display six representative power-law fits
reported by \citet{Adams2021}. The shapes of the multi-band LCs 
differ slightly depending on the {\bf $P_{\rm orb}$} used to fold the data
\citep[310--320\,days;][]{Casares2012,Moritani2015,Maier2019,Adams2021}.
We use folded LCs (Fig.~\ref{fig:fig10}) produced with the most recent orbital period
\citep[$P_{\rm orb}=317.3$\,days;][]{Adams2021}.
Note that the X-ray LC (Fig.~\ref{fig:fig10}, top) was constructed with average 0.3--10\,keV Swift-measured
fluxes \citep[first cycle and left ordinate;][]{Adams2021}
and average count rates (second cycle and right ordinate; TAH21);
the latter was normalized to have a maximum of 1.
The fluxes in the VHE LC \citep[Fig.~\ref{fig:fig10} bottom; taken from][]{Adams2021}
were obtained by assuming a photon index of
$\Gamma_{\rm VHE}=2.6$.

 	We base our broadband modeling of J0632 on the results of
the previous X-ray study (TAH21),
and use our model to explain the multi-band LCs and broadband SED of J0632.

\subsection{Application of the IBS model to J0632}\label{sec:sec4_2}

\begin{figure}
\centering
\hspace{7mm}
\includegraphics[width=73.1mm]{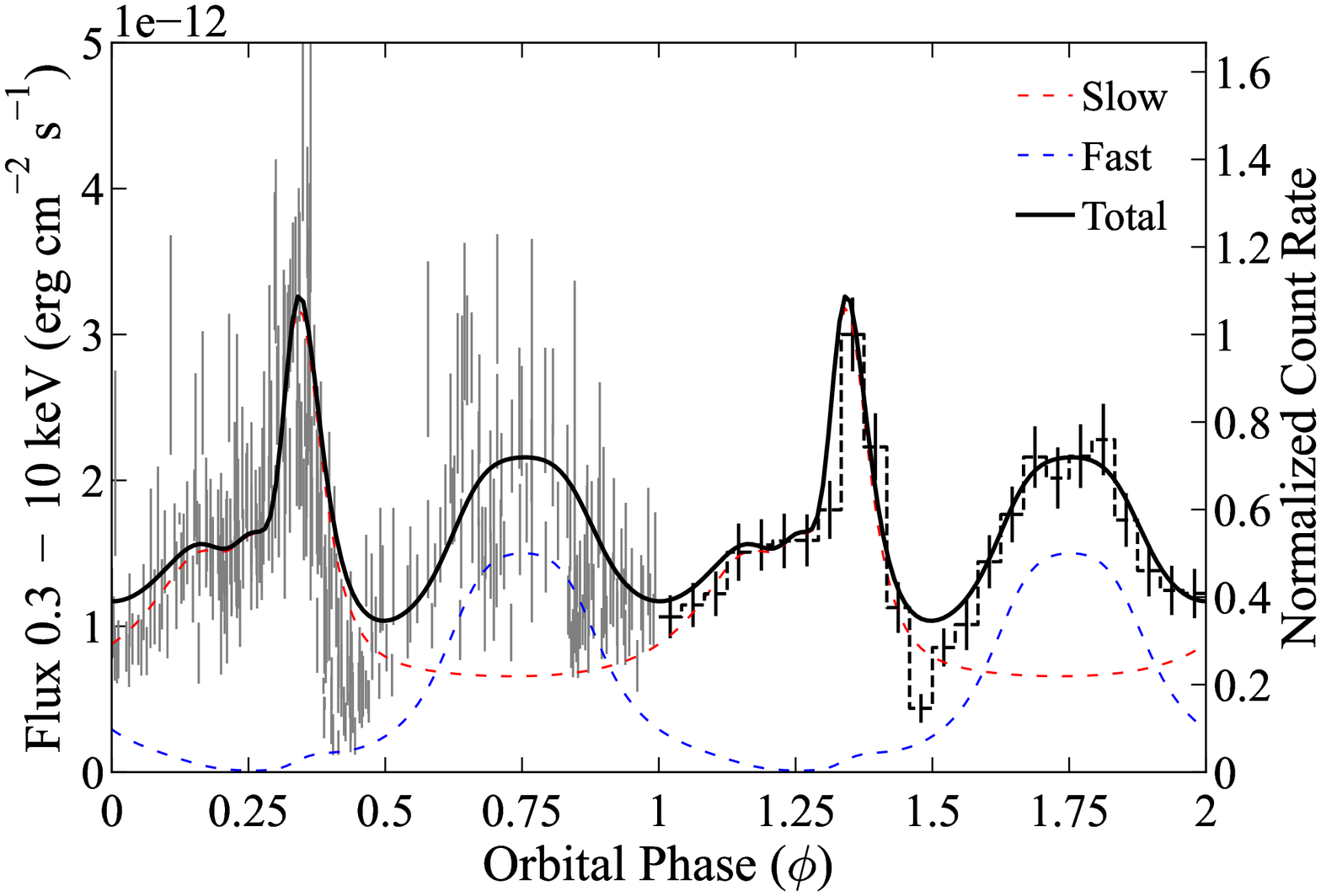}
\includegraphics[width=74 mm]{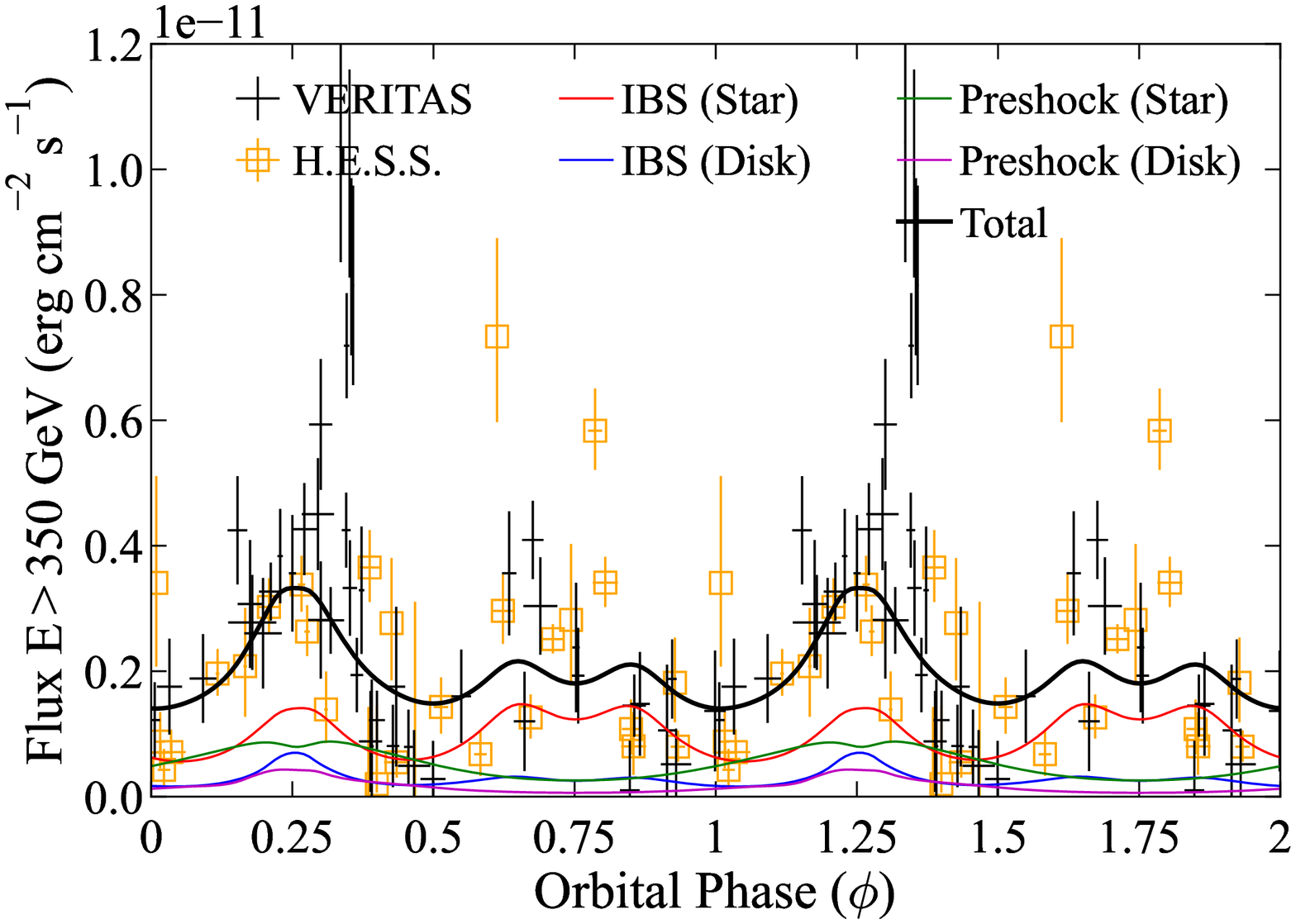}
\figcaption{Observed X-ray and VHE LCs of J0632.
{\it Top}: A Swift-measured 0.3--10\,keV LC of J0632 and an IBS model (black curve).
The fluxes in the first cycle (left ordinate) are taken from \citet{Adams2021}
and the normalized count rates in the second cycle (right ordinate) are taken from TAH21.
Red and blue dashed lines are model SY LCs of the slow and fast flows, respectively.
{\it Bottom}: $>$350\,GeV LCs measured by VERITAS (black) and H.E.S.S. (orange)
taken from \citet{Adams2021} and model computations.
EC emissions off of the stellar and disk photons by the IBS particles are shown in red and blue
solid lines, respectively, and the same by the preshock particles are presented
in green and purple, respectively. The sum of the EC emission components is
presented in black.
\label{fig:fig10}}
\end{figure}

	We apply our IBS emission model (Sections~\ref{sec:sec2} and \ref{sec:sec3})
to the observed SEDs and LCs of J0632.
We compute the IBS emission in 21 and 361 segments along $s$ and $\phi_s$, respectively, 
and the preshock EC emission in 500 radial segments. Particle spectra are constructed using
800 energy bins, and each of the resulting SY and EC emissions is computed in 300 frequency bins.
$\gamma$-$\gamma$ absorption
is applied to emission in each segment separately out to $\sim10^4d_{\rm orb}$ along the LoS,
and LCs are generated by integrating the computed SEDs in the energy ranges  
relevant to the X-ray (0.3--10\,keV) and VHE data ($>$350\,GeV).
The results are displayed in Figures~\ref{fig:fig9} and \ref{fig:fig10},
and the model parameters 
are presented in Table~\ref{ta:ta1}.

\begin{table}[t]
\vspace{-0.0in}
\begin{center}
\caption{Parameters for the IBS model in Figures~\ref{fig:fig9} and \ref{fig:fig10}}
\label{ta:ta1}
\vspace{-0.05in}
\scriptsize{
\begin{tabular}{lcc} \hline\hline
Parameter       & Symbol        & Value     \\ \hline
Semi-major axis & $a$ & $3.5\times10^{13}$\,cm \\
Eccentricity & $e$ & 0.45 \\
Inclination & $\theta_i$ & 48$^\circ$ \\
Periastron phase & $\phi_0$ & 0.26 \\
Pulsar inferior conjunction & $\phi_{\rm IFC}$ & 0.75 \\ \hline

Winds' momentum flux ratio & $\beta$ & 0.045 \\
Speed of slow flow at IFC & $v_{\rm flow}$ & $0.3c$ \\
Length of the IBS & $s_{\rm max}$ & $4d_{\rm orb}$ \\
Max. bulk Lorentz factor & $\Gamma_{\rm max}$ & 7 \\
Fraction of fast flow & $\xi$ & 0.05 \\
Magnetic-field strength$^{\rm a}$ & $B_0$ & 0.25\,G \\  \hline

Pulsar's power injection & $\eta \dot E_{\rm SD}$ & $2\times10^{34}\rm \ erg\ s^{-1}$ \\
Bulk Lorentz factor of preshock & $\gamma_{\rm e,peak}^{\rm pre}$ & $1.3\times 10^6$ \\
Min. electron Lorentz factor & $\gamma_{e,\rm min}$ & $1.5\times10^5$ \\
Max. electron Lorentz factor & $\gamma_{e,\rm max}$ & $\sim10^8$ \\
Lorentz factor at the break & $\gamma_b$ & $5\times10^6$ \\
Low-energy spectral index & $p_1$ & 2.3 \\
High-energy spectral index & $p_2$ & 2.6 \\ \hline

Radius of the companion & $R_*$ & 6.6$R_\odot$ \\ 
Temperature of the companion & $T_*$ & 30000\,K \\
Disk size & $R_{\rm d}$ & 60$R_*$ \\ 
Disk temperature & $T_{\rm d}$ & 0.7$T_*$ \\ \hline
\end{tabular}}
\end{center}
\vspace{-0.5 mm}
\footnotesize{
$^{\rm a}$ Magnetic-field strength at the apex ($s=0$) of the IBS at IFC.}
\end{table}

\subsubsection{Phase-averaged SED and LCs}\label{sec:sec4_2_1}
\begin{figure*}
\centering
\includegraphics[width=140mm]{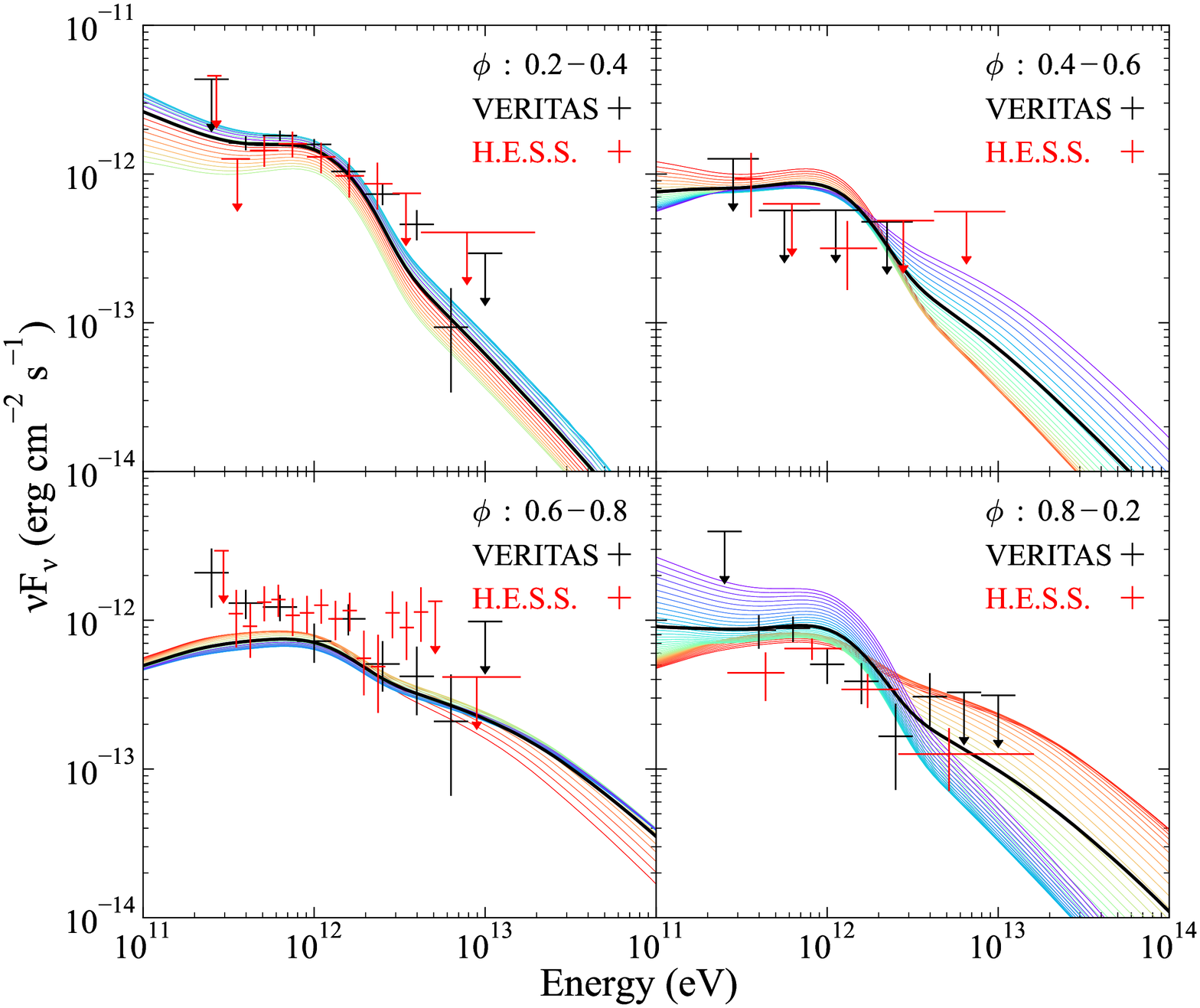}
\figcaption{Phase-resolved SEDs of J0632 \citep[][]{Adams2021} measured by VERITAS (black)
and H.E.S.S. (red). Model calculations at 20 phases within the designated phase interval are presented
in color: red to blue for early to late phases. The average of the computed SEDs in each interval
is displayed in black.
\label{fig:fig11}}
\end{figure*}

	The model reasonably describes the phase-averaged SED and the X-ray/VHE LCs
of J0632 (Figs.~\ref{fig:fig9} and \ref{fig:fig10}),
except for the spike at $\phi\approx0.35$ in the VHE LC
because we do not model the VHE emission at the disk crossing which requires deep 
understanding of how the pulsar and disk interact with each other. 
Notice that the model has a small bump at $\sim$TeV which is produced by the preshock-EC emission.
It has not been commonly included in the previous basic IBS models \citep[e.g.,][]{Dubus2015,An2017}, but
$\sim$TeV bumps seen in phase-resolved SEDs of J0632 (Fig.~\ref{fig:fig11}) hint at
a possibility of such preshock emission. By jointly modeling the X-ray and VHE data, we are able to infer the
magnetic-field strength $B_0$ (Eq.~\ref{eq:Bfield}) and
parameters for the particle spectrum (e.g., $\gamma_{e,\rm min}$
and $\gamma_{e,\rm max}$; Eq.~\ref{eq:bpl})
in the IBS since the spectral shapes of the IBS emission
depend strongly on these parameters.
The magnetic-field strength $B_0$ at the shock nose at IFC is inferred
to be $B_0\approx 0.25$\,G by the relative flux between the SY and the EC emissions, e.g.,
$\frac{F_{\rm SY}}{F_{\rm EC}} \approx \frac{u_{\rm B}}{u_*}$ in the Thomson regime for EC.

	For the $B_0$ value, $\gamma_{\rm e,min}$, $\gamma_{\rm e,max}$, and $\gamma_{\rm b}$
are inferred by the observed features in the SEDs since electrons with an energy $\gamma_e m_e c^2$
emit SY photons with an energy
\begin{equation}
\label{eq:syncnu}
h\nu_{\rm SY}\approx 4\times 10^6 h\left( \frac{B}{1\rm G} \right) \gamma_e^2\rm \ eV,
\end{equation}
where $h$ is the Planck constant, and upscatter seed photons with an energy $E_{\rm seed}$ to
\begin{equation}
\label{eq:ecnu}
h\nu_{\rm EC}\approx \gamma_e^2 \left( \frac{E_{\rm seed}}{1\rm eV}\right) \rm \ eV.
\end{equation}
$\gamma_{\rm e, min}$ is estimated to be $\approx10^5$ so that the low-energy cutoff in 
the model SED is below the 0.3\,keV start of the X-ray band
and the low-energy EC SED is below the LAT upper limits at $<$100\,GeV energies (Fig.~\ref{fig:fig9}).
$\gamma_{\rm e, max}$ is computed to be $\approx 10^{8}$ from the radiation reaction limit as mentioned in Section~\ref{sec:sec2_4}.
An upper bound for $\gamma_{\rm e, max}$ can be estimated since
a larger value (e.g., $\ge 4\times 10^{8}$) would overpredict the $<$GeV upper
limits. Constraints on the lower bound for $\gamma_{\rm e,max}$ are poor because
the VHE SED is little affected by $\gamma_{e,\rm max}$ due to the KN effect.
The observed $\sim$10\,TeV photons imply that $\gamma_{e,\rm max}>10^{6}$.
It is difficult to estimate $\gamma_{\rm b}$ without a measurement in the MeV band.

Our SY LC model for J0632 (Figure~\ref{fig:fig10}, top) fully matches the detected X-ray peak at $\phi\approx0.35$ and also produces an LC that is extremely similar to  that of TAH21, who phenomenologically modeled
the disk crossings as an enhancement of $B$ in the IBS. 
With this model, we ignore two complex phenomena - strong SY cooling and disk interactions - that impact the EC emission.  
In reality, SY cooling of particles would be boosted by the enhanced $B$.  However as our model 
does not include particle cooling, we are able to leverage the enhanced $B$ to fit the SY emission without altering the EC emission in the VHE band. 
Other processes such as additional seeds from disk heating may provide favorable conditions for EC emission
\citep{Chen2019}. As noted in Section~\ref{sec:sec2_1}, we also ignore these complexities in our model. 

The EC LC model (Figure~\ref{fig:fig10} bottom) reproduces two bumps:
one bump at $\phi\approx0.25$ produced by the slow flow from high seed density
and favorable ICS geometry at periastron and SUPC, and the other at $\phi\approx 0.75$  due to the Doppler boosted emission of the fast flow.
Note that there is a small notch at $\phi=0.75$ in the EC LC but not in the SY LC. This is due to the unfavorable
ICS scattering geometry at IFC ($\psi\approx 0$; see Section~\ref{sec:sec3_3}).
Although the notch cannot be identified in the data shown in the bottom panel of Figure~\ref{fig:fig10},
a smoothed LC presented in Figure~16 of \citet[][]{Adams2021} seems to exhibit
a notched morphology.

	We also display contributions of the stellar (red, green)
and disk (blue, purple) EC by the IBS and by the preshock particles
in the bottom panel of Figure~\ref{fig:fig10}.
Unlike the EC emission of the IBS particles, the preshock EC exhibits only
one bump at $\phi$=0.25 (near periastron and SUPC). Hence the preshock emission
controls the relative amplitude of the EC LC at $\phi$=0.25 and $\phi$=0.75;
using more preshock particles (e.g., larger $\eta \dot E_{\rm SD}$; Eq~\ref{eq:number})
and/or placing the emission zone (preshock acceleration region; Section~\ref{sec:sec2_3})
farther away from the pulsar would make the $\phi$=0.25 bump more pronounced.
As noted in Section~\ref{sec:sec3_3}, the EC off of the disk seeds is
weaker than that of the stellar seeds.
Although a distinction between the shapes of the stellar-EC
and the disk-EC emissions is subtle, the ratio of the emission peaks of the slow ($\phi=0.25$)
and the fast flow ($\phi=0.75$) particles is smaller for the stellar EC (red).
This is because the stellar EC at SUPC suffers from the KN effect, whereas
the IR seeds of the disk alleviates that effect. At IFC,
the small scattering angle $\psi$ and Doppler de-boosting
push the onset of the KN suppression to much higher $\gamma_e$s where little
or no scattering electrons are available, making the KN effect observationally
unimportant for both the stellar and disk seeds.

	The joint modeling of the X-ray and VHE LCs allows estimations of
the IBS size $s_{\rm max}$ ($\approx 4d_{\rm orb}$), $\Gamma_{\rm max}$ ($\approx 7$),
and $\xi$ ($\approx 0.05$)
because the ratio of the peak fluxes of the slow- and the fast-flow emissions
generated by the SY and EC processes
($F^{\rm slow, max}_{\rm SY}/F^{\rm fast, max}_{\rm SY}$
and $F^{\rm slow, max}_{\rm EC}/F^{\rm fast, max}_{\rm EC}$),
and the widths of the LC bumps at $\phi\approx 0.75$ depend
differently on these parameters (Figs.~\ref{fig:fig5} and \ref{fig:fig7}).
Note that the ratios $F^{\rm slow, max}_{\rm SY}/F^{\rm fast, max}_{\rm SY}$ and
$F^{\rm slow, max}_{\rm EC}/F^{\rm fast, max}_{\rm EC}$ are controlled by $\xi$, and
differences between them are determined by $s_{\rm max}$ and $\Gamma_{\rm max}$
(panels e and f in Figs.~\ref{fig:fig5} and \ref{fig:fig7}). The latter controls
also the widths of the LC peaks at IFC.

\subsubsection{Phase-resolved VHE SEDs}\label{sec:sec4_2_2}
	As the model reproduces the observed phase-averaged SED and multi-band LCs
reasonably well, we check to see if the model reproduces phase-resolved
VHE SEDs (Fig.~\ref{fig:fig11}).
Because the observed SED in a phase interval was constructed by combining 
observations from different epochs
with different exposures spread over the phase interval \citep[][]{Adams2021}, we display
model calculations at 20 phases within the interval (denoted in the panel) for comparison.

	The VERITAS- and H.E.S.S.-measured phase-resolved SEDs are reproduced well by
our model. Specifically, the modulation of the spectral hardness of the VHE spectra is mostly 
 consistent with the model; the hard VHE spectra at the phase interval 0.6--0.8 (Fig.~\ref{fig:fig11})
are explained by the enhanced contribution of the Doppler boosted emission of the fast flow.
Some noticeable discrepancies between the data and model are likely due to  
strong orbit-to-orbit variations of the VHE flux, as was noticed by \citet{Adams2021}, 
and uneven  phase coverage of the VHE measurements.

	It is intriguing to note that the observed SEDs
exhibit a bump at $\sim$TeV seen in phase intervals 0.2--0.4 and 0.8--0.2 (Fig.~\ref{fig:fig11}).
It was difficult to replicate this bump with basic IBS models which considered IBS particles only
as they produce a smooth SED. The preshock EC emission
(Section~\ref{sec:sec3_3} and Fig.~\ref{fig:fig9}) can naturally reproduce the SED feature,
which possibly suggests that the preshock emission is an important contributor to the VHE emission in J0632.

\section{Discussion\label{sec:discussion}}
In the previous sections, by adopting the latest orbital solution and binary system geometry suggested by a recent study (TAH21), we demonstrated that 
both the LC and SED data of J0632 from the X-ray to VHE  band can be well explained with our IBS model. Compared to the previous investigations based on X-ray data only, we further constrained the IBS parameters by
simultaneously fitting the X-ray and VHE data (Table~\ref{ta:ta1}). Still, we note that some of the parameters are degenerate, making error estimation difficult, and thus it is possible that the derived parameters may not represent a unique solution. In this section, we discuss several key IBS properties and observationally important parameters. 

\subsection{VHE emission} 
We found that most of the VHE flux in J0632 arises from EC 
of the stellar seeds by IBS electrons and that the exact shape of the seed spectrum (e.g., features
in the disk spectrum in Fig.~\ref{fig:fig3})
does not alter the EC spectrum significantly as long as
the IR-to-optical SED model matches the observed flux. For example, a phenomenological multi-temperature blackbody model
for the disk emission also results in a similar EC SED.
Notice that the disk-EC SED is slightly broader than the stellar-EC SED
(Section~\ref{sec:sec3_3} and Fig~\ref{fig:fig6}).
In particular, the former is less affected by the KN effect which is severe at
$h\nu_{\rm EC}\ge \frac{(m_e c^2)^2}{E_{\rm seed}} \rm \ eV$, hence making 
the SED model spectrally harder if the intrinsic energy density of the disk emission
is much higher. The observationally inferred $u_{\rm *}$ of the disk depends
on the assumed disk inclination $i_{\rm d}$ with respect to the LoS.
We used an extreme inclination angle of $i_{\rm d}=85^\circ$  for modeling the disk emission (Section~\ref{sec:sec3_1});
if this value were smaller, the intrinsic disk emission would be lower, thus 
affecting the total observed VHE emission less significantly.

The `total' disk + star seed flux inferred from the optical data (Fig.~\ref{fig:fig3})
is an important factor that determines the amplitude of the VHE emission.
The inferred optical seed density $u_{\rm *}$ in the emission zones
depends on the assumed distance to the source (here, $d$=1.4\,kpc)
and orbital size quantified by the semi-major axis $a$.
Hence, these parameters $d$ and $a$ have large impacts on our IBS parameter determination because one of the most important parameters $B_0$ is related to $u_{\rm *}$ (Section~\ref{sec:sec4_2_1}). 
Our estimations of $B_0$ and thus $\gamma_{\rm e,min}$ and $\gamma_{\rm e,max}$
are sensitive to the value of 
$u_{\rm *}$; see Eqs.~\ref{eq:syncnu} and \ref{eq:ecnu}. Therefore, accurate measurements of $d$ and $a$ as well as precise characterizations of the X-ray and VHE emission will lead to better understanding of the particle properties and thus particle acceleration 
in the IBS.

\subsection{Particle flow in the IBS region} 
We were able to infer $s_{\rm max}\approx 4d_{\rm orb}$,
$\Gamma_{\rm max}\approx 7$, and $\xi\approx 0.05$ by modeling the
X-ray and VHE LCs. Since the quality of the VHE LCs is rather poor, 
our estimations are subject to large uncertainties. 
Moreover, model parameters are covariant with each other,
and our simplified prescriptions such as the IBS shape and $B$ structure 
for the IBS flows may not be very accurate.
These may add further systematic uncertainties. 
Although the reported parameter values (Table~\ref{ta:ta1}) need to be taken with some caution,
we point out that the inferred parameters are in accord with the recent HD simulations
supporting the general properties of IBS flows in TGBs -- 
high-energy emission arises from the inner region of the IBS flow and
a small fraction of the particles are bulk accelerated in the flow.

\subsection{A TeV bump in the SED: EC by preshock particles} 
The most interesting emission feature in J0632 is the $\sim$TeV SED bump seen
in some orbital phases, which we ascribed to EC emission of the preshock particles
with $\gamma^{\rm pre}_{\rm e,peak}\sim 10^6$. It has been widely believed that
pulsar wind is dominated by Poynting flux near the magnetosphere, and 
the magnetic energy should be converted into particle energy 
in the pulsar wind zone between the light cylinder and the termination shock.
The location, physical processes, and energetics for such an energy conversion have
not been well understood theoretically \citep[e.g.,][]{Coroniti1990,Cerutti2020}.
But \citet{Aharonian2012} modeled pulsed VHE emission of the Crab nebula and
suggested that the Poynting flux of the pulsar wind should be converted into kinetic energy of particles in a narrow region at
$\approx$30$R_{\rm LC}$, and that the accelerated particles have a Lorentz factor
of $\gamma^{\rm pre}_{\rm e,peak}\sim 10^6$.

The $\sim$TeV bump in J0632's SEDs
can be explained by such preshock emission as we demonstrated in Section~\ref{sec:sec4_2_2}. 
Note that similar SED features have been seen in the VHE low state of the TGB
PSR~J2032+4127 \citep[e.g.,][]{VeritasJ20322018}.
The relatively sharp $\sim$TeV bumps imply that
the energy distribution of the preshock is narrow like
the Maxwellian distribution we assumed (Eq.~\ref{eq:maxwellian}).
Other narrow distributions such as a broadened delta function or a narrow power law
with a peak at $\gamma^{\rm pre}_{\rm e,peak}\sim10^6$
would predict slightly different SED shapes for the bumps but
may also explain them equally well. Currently, these distributions are indiscernible because the measurement uncertainties are large. 
Precise characterization of the bumps with deep VHE observations may
help to constrain the preshock particle distribution, possibly providing a hint to
acceleration mechanisms \citep[e.g.,][]{Hoshino1992,Jaroschek2008,Sironi2011} in the pulsar wind zone.

In our model, the SED amplitude, emission  frequency (Fig.~\ref{fig:fig11}),
and the LC shape (Fig.~\ref{fig:fig10} bottom) of the preshock EC 
depend sensitively on the conversion location and particle energy; 
if the conversion takes place at a larger distance from the pulsar, the model predicts weaker preshock emission and sharper LC features (particularly, a sharper peak near periastron is expected). Thus,  the SED features in TGBs 
can provide a sensitive probe to the energy conversion process
in pulsar winds. While the current measurements of the SED and LC of J0632 are somewhat constraining,
more accurate measurements with deeper VHE observations \citep[e.g., by CTA;][]{CTA2011}
in the near future will enable us to determine the fundamental parameters for energy conversion
in pulsar winds.

\subsection{A sharp spike in the VHE LC: a signature of disk-pulsar interaction? } 
Since the large spike observed at $\phi = 0.35$ in the VHE LC was not accounted for in our baseline IBS model, we suspect that this unique feature  may be caused by a disk crossing of the pulsar.
Note that the circumstellar disk material was suggested to be the origin of the higher $N_{\rm H}$ at the disk interaction phases  \citep[TAH21;][]{Malyshev2019}. If this is true, the disk interaction may leave some observable signatures in the multi-band emission at the `interaction' phases.

\citet[][]{HESS2020} reported an increase in the VHE flux near the disk crossing phase
for another TGB PSR~B1259$-$63,
and \citet{Chen2019} attributed the enhanced VHE flux to disk heating by the IBS. 
Similarly, we find that the VHE spike in the LC can be reproduced if we assume that the IBS is immersed in a 
$T_{\rm BB}\approx 950$\,K blackbody field of a heated disk.
Note, however, that the existence of $T_{\rm BB}\approx 950$\,K blackbody field is
not physically supported by our model since it does not include complex dynamic effects and
microphysics of interaction such as disk heating. We defer further investigations to a future work.

\subsection{Comparison with other IBS models: double-bump features in the LCs} 
A difference between the IBS model (TAH21) and the similar inclined disk
model for J0632 \citep[][]{Chen2022} is that the latter assumed one-zone shock geometry 
whereas the former used an extended cone-shape IBS with two particle flows.
This difference has a significant impact on the LC modeling. The one-zone shock model
requires a peculiar ``inclined disk'' geometry to account for the two bumps in the LC \citep{Chen2022}.
In contrast, the IBS cone with two particle flows can naturally reproduce two bumps, and disk crossing can
add two more (bumps or peaks). Overall, the IBS model was able to accommodate the complex X-ray
LC of J0632 with two bumps and one peak (TAH21).

	The most distinctive feature of IBS models, as compared to one-zone
inclined disk models, is a double-peak structure
around IFC in the X-ray/VHE LCs of highly
inclined sources
\citep[Figs.~\ref{fig:fig5} and \ref{fig:fig7}; see also][]{vandermerwe2020}.
Such double-peak features are often observed in the X-ray LCs
of redback pulsar binaries and regarded as a signature
of bulk-accelerated particles in the IBS \citep[e.g.,][]{Kandel2019}.
Intriguingly, a hint of double-peak X-ray/VHE LCs was also  found in J0632 \citep[Figs.~15 and 16 of][]{Adams2021}, suggesting that the emission at IFC is indeed produced by IBS cone emission. 
While our SY model based on the parameters derived in TAH21 does
not reproduce the X-ray double peaks at $\phi=0.75$, the IBS model can account for 
the double-peak LC with a larger inclination (e.g., Fig.~\ref{fig:fig5} c); in this case,
the model-predicted dip at $\phi=0.75$ in the EC LC (Fig.~\ref{fig:fig10} bottom)
might be deeper. This prediction can be confirmed with more sensitive X-ray and VHE observations
and can distinguish between the disk interaction case  \citep[][]{Chen2022}
and the IBS cone emission.

\subsection{The origin of short-term variability} 
	Broadband emission of J0632 is known to exhibit strong orbit-to-orbit
variability \citep[TAH21;][]{Adams2021}. In the IBS scenario, such short-term variability is likely due to non-uniform clumpy stellar winds that can drive varying momentum flux ratio $\beta$. In this case, the shock opening angle $\theta_{\rm cone}$
\citep[Eq.~\ref{eq:coneangle}; see also][]{Bogovalov2008}, and distances to the emission zones from the pulsar $r_{\rm p}$ and the star $r_{\rm s}$
are also expected to vary, affecting the IBS emission.
For example, 
a stronger stellar outflow will push the IBS closer to the pulsar, making $B$ larger by reducing $r_{\rm p}$
and $u_{\rm *}$ smaller by increasing $r_{\rm s}$. This will enhance the SY emission but 
reduce the EC emission in general as seen in panel d of Figs.~\ref{fig:fig5} and \ref{fig:fig7}.

At the IFC, however, the situation is more complicated because 
the Doppler factor $\delta_{\rm D}$, which is also determined by $\beta$ through
a change in $\theta_{\rm cone}$,
comes into play. Generally, stronger variability is expected at the IFC if the LoS 
is close to the shock tangent as it is in J0632, because the observed flux depends strongly on $\delta_{\rm D}$ \citep[e.g.,][]{An2017}. 
Contemporaneous observations of multi-band variability in the optical,
X-ray, and VHE bands will provide a useful diagnostic to test the IBS scenario.

\section{Summary\label{sec:summary}}
We showed that our phenomenological IBS model provides a good fit to the multi-band SED and LC data of J0632, 
suggesting that the interaction between the winds of the pulsar and companion drives the observed broadband emission. Below we summarized our results and conclusions. 

\begin{itemize}
\item We constructed an IBS emission model employing physical
processes and emission components appropriate for TGBs,
applied the model to J0632 with an orbit inferred from
an X-ray LC modeling (TAH21), and found that
the model and the orbit could explain not only the X-ray LC but also
the VHE emission properties of the source LCs and SEDs.

\item The observed SEDs of J0632 show a
bump at $\sim$TeV in some phase intervals. 
This feature is likely due to ICS emission by preshock
particles, implying that the bulk Lorentz factor
of the preshock is $\gamma^{\rm pre}_{\rm e,peak}\sim 10^6$. 
Conversely, accurate characterizations of the bump, both observationally and theoretically, will help us to understand the energy conversion processes in pulsar winds.

\item Our VHE LC model for J0632 predicts a double-peak structure around IFC. This is a natural consequence of the IBS model in contrast to inclined disk models \citep[e.g.,][]{Chen2022}. 
\end{itemize}

	While the model captures the main features of the broadband emission SED and LCs of J0632, some of the parameters such as
$\gamma_b$ are not well constrained. Further, the parameters are degenerate and thus it was difficult to determine 
a unique solution set. 
A comparison with MHD simulations as well as future contemporaneous observations in the optical, X-ray, and VHE bands will allow us to determine these parameters better and to break the degeneracy. 
Further constraints on the IBS properties can be set by detecting a spectral break caused by particle cooling. 
The break is expected at $\sim$MeV energies where future missions \citep[e.g., COSI;][]{Tomsick2019} can add valuable data in the MeV band to boost  our understanding of TGBs.

It is well known that less powerful IBSs are formed in the so-called `black widow' and `redback' pulsar binaries \citep[e.g.,][]{Romani2016,Wadiasingh2017}.
The common signatures expected from IBSs formed in pulsar binaries are double-peak X-ray LCs (for large $\theta_i$) and orbital modulations in the $\sim$GeV band \citep[e.g.,][]{An2018,Corbet2022}.
Hence, a variety of IBS models, including ours  presented in this paper, has been applied to pulsar binaries in circular orbits with a very low-mass companion \citep[e.g.,][]{vandermerwe2020}.
However, it is yet unclear whether an (universal) IBS model  can eventually account for the diverse
observational properties of both the pulsar binaries and the TGBs by simply
employing different geometries and energetics; e.g., the X-ray, GeV, and VHE phase variations
are observed to be diverse in the sources \citep[e.g.,][]{Corbet2012,Corbet2016,An2020}.
While the pulsar binary and TGB systems certainly share some common emission properties (e.g., orbital modulation), the systems seem to possess fundamentally different mechanisms  (e.g., companion, orbit, energetics etc). A large and more comprehensive multi-wavelength study of these exotic pulsar binaries and TGBs, as demonstrated for J0632 in this paper, will give deeper insights into IBS and pulsar physics.

\acknowledgments
We thank Melania Nynka for a helpful review of the paper.
We thank the anonymous referee for the careful reading of the paper and insightful comments.
This research was supported by Basic Science Research Program through
the National Research Foundation of Korea (NRF)
funded by the Ministry of Science, ICT \& Future Planning (NRF-2022R1F1A1063468).

\bigskip
\vspace{5mm}

\bigskip
\bibliographystyle{apj}
\bibliography{ms}

\begin{appendix}
	The shape of an IBS formed by interaction of isotropic winds of two stars was
analytically calculated by \citet{Canto1996}, and computations of SEDs of the SY and ICS emissions
were well described in \citet{Finke2008} and \citet{Dermer2009}. Here we show formulas
that we used for the IBS emission model for reference.

\section{Formulas for computation of the IBS shape}
\label{sec:appendix1}
	Interaction of two isotropic winds forms a contact discontinuity (CD) of which shape
can be analytically computed using pressure balance equations \citep[][]{Canto1996}.
For a wind's momentum flux ratio of $\beta$ (Eq.~\ref{eq:beta})
and the geometry depicted in Figure~\ref{fig:fig2}, the locus of the CD in a vertical plane
(to the orbital plane) containing the pulsar and the star is given by
\begin{equation}
\label{eq:CDrad}
r_{\rm p} = d_{\rm orb} \mathrm{sin}\theta_{\rm s} \mathrm{csc}(\theta_{\rm p} + \theta_{\rm s}),
\end{equation}
and
\begin{equation}
\label{eq:CD}
\theta_{\rm s} \mathrm{cot}\theta_{\rm s} = 1 + \beta(\theta_{\rm p}\mathrm{cot}\theta_{\rm p} - 1)
\end{equation}
\citep[Eqs.~23 and 24 of][]{Canto1996}.
Distance to the apex of the CD ($\theta_{\rm s}=\theta_{\rm p}=0$; shock nose) from the pulsar
and the asymptotic angle of the IBS
($\theta_{\rm cone}$; half opening angle of the IBS cone) are given by
\begin{equation}
\label{eq:shocknose}
r_0 = d_{\rm orb}\frac{\sqrt{\beta}}{1+\sqrt{\beta}}
\end{equation}
and
\begin{equation}
\label{eq:coneangle}
\mathrm{tan}\theta_{\rm cone} - \theta_{\rm cone} = \frac{\pi \beta}{1-\beta},
\end{equation}
respectively \citep[Eqs.~27 and 28 of][]{Canto1996}.
$\theta_{\rm cone}$ increases monotonically with increasing $\beta$.

\section{Formulas for computation of emission SEDs}
\label{sec:appendix3}
	Suppose particle flow has bulk motion with a Lorentz factor $\Gamma$ with respect
to an observer, and that the particles in the flow-rest frame move randomly (isotropically) with 
an energy distribution $\frac{dN_e'(\gamma')}{d\gamma'}$, where $\gamma'$ is the Lorentz factor
of the particle's random motion in the flow rest frame (primed quantities are defined in the
flow rest frame). In randomly oriented $B$, the SY SED of the particles can be
computed using the formula given in Eq.~(18) of \citet{Finke2008};
\begin{equation}
\label{eq:sysed}
f_{\rm SY}(\epsilon)=\frac{\sqrt{3}\delta_{\rm D}^4\epsilon'q_e^3B}{4\pi hd^2}\int_1^\infty d\gamma'
\frac{dN_e'(\gamma')}{d\gamma'}R(x),
\end{equation}
and
\begin{equation}
\label{eq:Rx}
R(x) = \frac{x}{2}\int_0^{\pi} d\theta \mathrm{sin}\theta \int_{x/\mathrm{sin}\theta}^{\infty} dt K_{5/3}(t)
{\rm \ with} \ x=\frac{4\pi \epsilon' m_e^2 c^3}{3q_e B h\gamma'^2},
\end{equation}
where $\epsilon=h\nu/m_ec^2$ and $\epsilon'=h\nu'/m_e c^2$ are
dimensionless energies of the observed (observer frame) and the emitted (flow-rest frame) photons, respectively,
$m_e$ is the mass of an electron, $q_e$ is the electron charge,
$h$ is the Planck constant, $d$ is the distance between the emission zone and the observer,
$K_{5/3}$ is a modified Bessel function,
and $\delta_{\rm D}$ is a Doppler beaming factor determined by the flow viewing angle $\theta_{\rm V}$
(angle between flow tangent and observer) and the bulk Lorentz factor $\Gamma$:
\begin{equation}
\label{eq:Doppler}
\delta_{\rm D}=\frac{1}{\Gamma(1 - \sqrt{1 - 1/\Gamma^2}\mathrm{cos}\theta_{\rm V})}.
\end{equation}

        Energetic particles in the preshock and IBS can upscatter ambient low-energy
photons emitted by the star (blackbody) and the disk via the ICS process (i.e., EC).
The scattering cross section depends on the scattering geometry \citep[e.g.,][]{Dubus2013}
and can be simplified with a head-on approximation if the electrons are highly relativistic
\citep[i.e., head-on collision;][]{Dermer2009}.
Because the electrons in the IBS and preshock of TGBs are highly relativistic
(i.e., $\gamma_{\rm e}\gg1$; Sections~\ref{sec:sec2_3}--\ref{sec:sec2_4}),
we use the head-on approximation and calculate EC SEDs using Eq.~(34) of \citet{Dermer2009}:
\begin{equation}
\label{eq:ecsed}
f_{\rm EC}(\epsilon_s)=\frac{3c\sigma_{\rm T}}{32\pi d^2}\epsilon_{\rm s}^2 \delta_{\rm D}^3\int_0^{2\pi}d\phi_*
\int_{-1}^1 d\mu_*
\int_0^{\epsilon_{*,\rm hi}} d\epsilon_* \frac{u_*(\epsilon_*,\Omega_*)}{\epsilon_*^2}
\int_{\gamma_{\rm low}}^{\infty} d\gamma \frac{dN'_e(\gamma / \delta_{\rm D}) }{d\gamma}\frac{\Xi}{\gamma^2},
\end{equation}
where $\sigma_{\rm T}$ is the Thomson scattering cross section, $\epsilon_*$ and $\epsilon_s$
are dimensionless energies ($\frac{h\nu}{m_e c^2}$) of the incident and scattered photons, $\phi_*$ and $\mu_*$ are the
direction of the incident photon into the emission zone,
$u_*(\epsilon_*, \Omega_*)$ is the energy density of the seed photons in the emission zone, and
$\Xi$ is defined as 
\begin{equation}
\label{eq:Xi}
\Xi \equiv y + y^{-1} - \frac{2\epsilon_s}{y\bar \epsilon y} +\left (\frac{2\epsilon_S}{y\bar \epsilon y} \right )^2
{\rm\ with\ } y\equiv 1-\frac{\epsilon_s}{\gamma},
\end{equation}
where $\bar \epsilon$ is the invariant collision energy:
\begin{equation}
\label{eq:ebar}
\bar \epsilon \equiv \gamma \epsilon_*(1-\sqrt{1-1/\gamma^2}\mathrm{cos}\psi)
\end{equation}
with $\psi$ being the scattering angle of the incident photon.
The integration limits in Eq.~\ref{eq:ecsed} are determined by the scattering kinematics and are
\begin{equation}
\label{eq:gammalow}
\gamma_{\rm low} = \frac{\epsilon_s}{2}\left[ 1 + \sqrt{1 + \frac{2}{\epsilon_* \epsilon_s(1-\mathrm{cos}\psi)}} \right ],
\end{equation}
and
\begin{equation}
\label{eq:ehi}
\epsilon_{*,\rm hi}= \frac{2\epsilon_s}{1-\mathrm{cos}\psi}.
\end{equation}

\end{appendix}
\end{document}